%% file: 00_main.tex
\documentclass[12pt, draftclsnofoot, onecolumn]{IEEEtran}

\usepackage[T1]{fontenc}
\usepackage{siunitx}
\usepackage{cite}
\usepackage[cmex10]{amsmath}
\usepackage{amssymb}
\usepackage{makecell}
\usepackage{boldline}
\usepackage{adjustbox}
\usepackage{caption}
\usepackage{tikz}
\usetikzlibrary{shapes,arrows,fit,positioning,shadows,calc}
\usetikzlibrary{plotmarks}
\usetikzlibrary{decorations.pathreplacing}
\usetikzlibrary{patterns}
\usepackage{algorithm}
\usepackage{algpseudocode}
\algnewcommand{\Initialization}[1]{%
  \State \textbf{initialization:}
  \Statex \hspace*{\algorithmicindent}\parbox[t]{.8\linewidth}{\raggedright #1}
}
\usetikzlibrary{automata}
\usepackage{xcolor}

\usepackage{pgfplots}
\pgfplotsset{compat=newest}

\usepackage{graphicx}
\usepackage{ragged2e}
\usepackage{hhline}

\newcommand{\pr}[1]{\ensuremath{\left[#1\right]}}
\newcommand{\pc}[1]{\ensuremath{\left(#1\right)}}
\newcommand{\chav}[1]{\ensuremath{\left\{#1\right\}}}
\newcommand{\PM}[1]{\ensuremath{\left|#1\right|}}

\definecolor{b}{rgb}{0, 0, 1}
\definecolor{r}{rgb}{1, 0, 0}
\definecolor{r2}{rgb}{0,0,0}

\definecolor{dark_green}{rgb}{0, 0.33, 0.13}
\definecolor{naplesyellow}{rgb}{0.99, 0.93, 0.0}
\definecolor{aureolin}{rgb}{1, 0.8, 0}
\definecolor{purple}{rgb}{0.4940 0.1840 0.5560}

\ifCLASSINFOpdf
\else
\fi
\usepackage{amsmath}
\usepackage[cmintegrals]{newtxmath}
\begin{document}
%

%
\title{Minimum Symbol Error Probability Low-Resolution Precoding for MU-MIMO Systems With PSK Modulation}

%
%

\author{Erico~S.~P.~Lopes,~\IEEEmembership{Graduate Student Member,~IEEE},
				~{Lukas~T.~N.~Landau,~\IEEEmembership{Member,~IEEE},~and~{Amine~Mezghani,~\IEEEmembership{Member,~IEEE}}}
\thanks{E.~S.~P.~Lopes and L.~T.~N.~Landau are with the Centro de Estudos em Telecomunica\c{c}\~{o}es, Pontif\'{i}cia Universidade Cat\'{o}lica do Rio de Janeiro, Rio de Janeiro CEP 22453-900, Brazil, (e-mail: lukas.landau@cetuc.puc-rio.br; erico@cetuc.puc-rio.br).
A.~Mezghani
 is with the Department of Electrical and Computer Engineering, University of Manitoba, Winnipeg, MB,
R3T 5V6, Canada, (e-mail:
amine.mezghani@umanitoba.ca).
This work has been supported by the {ELIOT ANR18-CE40-0030 and FAPESP 2018/12579-7} project.
}
}
\maketitle


\begin{abstract}
We propose an optimal low-resolution precoding technique that minimizes the symbol error probability of the users. Unlike existing approaches that rely on QPSK modulation, for the derivation of the minimum symbol error probability objective function the current approach allows for any PSK modulation order. Moreover, the proposed method solves the corresponding discrete optimization problem optimally via a sophisticated branch-and-bound method.
Moreover, we propose different approaches based on the greedy search method to compute practical solutions. 
Numerical simulations confirm the superiority of the proposed minimum symbol error probability criteria in terms of symbol error rate when compared with the established MMDDT and MMSE approaches.
\end{abstract}

\begin{IEEEkeywords}
Precoding, Constant Envelope, Low-Resolution Quantization, Phase Quantization, MIMO systems, Minimum Symbol Error Probability, Branch-and-Bound methods, Greedy Search Algorithms, Projection Based methods.
\end{IEEEkeywords}


%
\IEEEpeerreviewmaketitle

\input{01_Introduction}
\input{02_System_Model}
\input{Formulations/Formulations}

\input{Precoder/01_Precoding}

\input{Numerical_results/01_Numerical_Results}

\input{03_Conclusion}

\appendix

\input{Appendixes/Appendix_Precoder}

\ifCLASSOPTIONcaptionsoff
  \newpage
\fi

\bibliographystyle{IEEEtran}
\bibliography{bib-refs}
\end{document}

%% file: 01_Introduction.tex
\section{{Introduction}}

\IEEEPARstart{M}ultiuser multiple-input multiple-output (MU-MIMO) systems are considered as a promising physical-layer technique and are expected to be vital for future of wireless communications networks \cite{6G_Future_Directions}. 
However, due to the high number of radio frequency front ends (RFFE) the energy consumption and hardware costs of the radio frequency chains impose a challenge for this kind of technology \cite{Power_consumption}.

According to \cite{6G_100GHz} energy efficiency (EE) will be a key feature of the next generation wireless communications networks. As stated in \cite{6G_Use_cases}, 6G networks will require 10 to 100 times higher EE when compared to 5G, to enable scalable low-cost deployments, with low environmental impact, and better coverage. Another central demand for the future of wireless communications is higher data reliability, \cite{6G_research_challanges,6G_Vision}. As such, one challenge for MU-MIMO systems design is the increase in EE with minimum the bit-error-rate (BER) compromise.

To maximize the EE of MIMO systems, different studies, e.g. \cite{Power_efficiency} and \cite{min_power_cons}, have been conducted on how to decrease the energy consumption of a RFFE. One of the main approaches present in literature is to consider low-resolution data converters and constant envelope signaling. The main drawback of adopting these features is the performance degradation in the BER they yield.

\subsection{Related Works}

To mitigate the performance degradation low-resolution precoding have been receiving increasing attention of the wireless communications community. 
Several precoding strategies with low-resolution data converters exist in literature. Linear approaches, such as the phase Zero-Forcing (ZF-P) precoder \cite{ZF-Precoding}, the Wiener Filter Quantized (WFQ) precoder \cite{Jacobsson_2017} and the Sparse-ZF method \cite{Amine_SSP_2021}, have been proposed and benefit from a relatively low computational complexity.
However, to achieve a higher degree of reliability nonlinear symbol level precoding (SLP) methods have been presented based on different design criteria. The most popular criteria are the minimum mean squared error (MMSE) and the maximum minimum distance to the decision threshold (MMDDT). 

For MMSE different SLP strategies have been presented. In \cite{Squid_precoder} a 1-bit MMSE precoder was devised based on the semidefinite relaxation of the discrete feasible set. In \cite{jacobsson2018nonlinear} another MMSE based approach was proposed, this time employing a relaxation based on the infinity-norm for MU-MIMO-OFDM systems. In \cite{Jedda_mmse_mapped} a modified MMSE objective was considered and the feasible set was relaxed to its convex hull of a M-PSK modulation. In \cite{Partial_BnB} a partial branch and bound (B\&B) algorithm was considered in the context of QAM signaling for the 1-bit case. Using the MMSE criterion some optimal precoding approaches have also been developed. In \cite{Jacobsson2018} a sphere precoding strategy is utilized to compute the MMSE optimal transmit vector for the 1-bit case. Moreover, the studies from \cite{MMSE_bb} and \cite{lopes2021discrete} proposed different B\&B techniques that yield the optimal MMSE precoding vector for any M-PSK modulation.

The MMDDT criterion, also called maximum safety margin (MSM) criterion or constructive interference (CI) criterion, have been receiving great attention of the community. In \cite{CVX-CIO} different a constant envelope designs have been proposed based on the MMDDT criterion under per-antenna-power constraint. In \cite{MSM_precoder} the MMDDT criterion is also utilized this time under the convex hull constraint. In \cite{Nossek_GS_mmddt} and \cite{Hong_GS_mmddt_PSK} different greedy-search (GS) algorithms were developed using the MMDDT criterion for the 1-bit case with PSK modulation. The work from \cite{Hong_GS_mmddt_PSK} is extended in \cite{Hong_GS_mmddt_qam} for QAM modulation. In \cite{Partial_BnB} a partial B\&B algorithm was considered in the context of PSK signaling for the 1-bit case. In \cite{Landau2017} a B\&B technique is utilized to compute the optimal MMDDT precoding vector also for the 1-bit case. Finally, the study from \cite{General_MMDDT_BB} generalizes the work from \cite{Landau2017} for phase quantization with an arbitrary number of bits.

The study from \cite{lopes2021discrete} provides some useful insights about the relation between MMSE and MMDDT. Based on optimal transmit vectors \cite{lopes2021discrete} shown that the MMSE outperforms MMDDT in terms of BER for low-SNR regime, while the opposite is true for the high-SNR range. Different works, e.g. \cite{General_MMDDT_BB} and \cite{Jedda_2016}, claim that the MMDDT criterion performs optimally for the high-SNR regime in terms of symbol error rate (SER).

Aside from the MMSE and MMDDT criteria some works consider the direct optimization of symbol error probability (SEP) for precoding design e.g. \cite{Mingjie_framework,Sohrabi2018,Mingjie_icassp2018,Mingjie_SSP2018,Mingjie_globalsip2018,mezghani2020massive}. While the studies from \cite{Mingjie_framework,Sohrabi2018, Mingjie_icassp2018,Mingjie_SSP2018} mainly focus on QAM signaling with 1-bit data converters, \cite{Mingjie_globalsip2018} considers M-PSK modulation shows that by using the MMDDT design one can minimize an upper bound on the SEP. Another relevant result was achieved in \cite{mezghani2020massive} where an analytical formula for the SEP using QPSK data modulation is derived and, based on it, a discrete optimization problem is formulated. 

\subsection{Main Contributions}

Following the direction from the works from \cite{Mingjie_framework,Sohrabi2018,Mingjie_icassp2018,Mingjie_SSP2018,Mingjie_globalsip2018,mezghani2020massive} this study considers the direct minimization of the SEP (MSEP) criterion and focuses on the development of precoding techniques for a MU-MIMO downlink system with PSK modulation and phase quantization with an arbitrary number of bits.

In this context, we consider for the design objective the minimization of the exact SEP presented by \cite{mezghani2020massive} as the design criterion for the QPSK case and derive a novel MSEP formulation based on the union bound probability for higher-order PSK cases. Note that, the proposed union-bound MSEP (UBMSEP) formulation, similarly as in the MMDDT case, relies on the minimization of an upper bound of the SEP.
Numerical results confirm that the MSEP formulations outperform, in terms of SER, the MMSE criterion for medium and high SNR regime, while having similar performance for low-SNR, and outperform, in terms of SER, the MMDDT criterion for all examined SNR range.

The mentioned MSEP formulations are then utilized in conjunction with different optimization techniques to develop diverse precoding algorithms. First, projection based methods (PBMs) are considered as practical precoding techniques. PBM is a popular class of approaches present in literature, many works have utilized PBMs to devise practical precoding algorithms, e.g. \cite{Hong_GS_mmddt_PSK,Nossek_GS_mmddt,lopes2021discrete,Jedda_mmse_mapped,MSM_precoder,MSM_fista}. Then MSEP B\&B approaches are considered for the computation of the optimal precoding vector. The proposed B\&B algorithms differ from the ones present in literature, e.g. \cite{Landau2017,General_MMDDT_BB,MMSE_bb,lopes2021discrete}, in terms of using a more restrictive pruning step which, then, contributes to a decreased average number of evaluated bounds and also in terms of the flexibility for choosing the projection step.

\subsection{Remainder and Notation}

The remainder of this paper is organized as follows: Section~\ref{sec:system_model} describes the system model, whereas Section~\ref{sec:precoding_formulation} exposes the derivation of the precoding design criteria. Section~\ref{sec:PBM_MSEP} devises practial projection based precoding methods based on the presented formulations. Section~\ref{sec:MSEP_BB} derives a B\&B algorithm to compute the optimal MSEP solution. 
Section~\ref{sec:numerical_results} presents and discusses numerical results, while Section \ref{sec:conclusion} gives the conclusions. A convexity analysis is provided in the appendix.

Regarding the notation, bold lower case and upper case letters indicate vectors and matrices respectively. Non-bold letters express scalars. The operators $(\cdot)^*$ and $(\cdot)^T$ denote complex conjugation, transposition and Hermitian transposition, respectively. Real and imaginary part operator, as well as the functions $\Phi(\cdot)$, $\text{Q}(\cdot)$, $\text{erf}(\cdot)$  and $\log(\cdot)$, are also applied to vectors and matrices, e.g., $\mathrm{Re}\left\{ \boldsymbol{x} \right\} = \left[ \mathrm{Re}\left\{ \left[\boldsymbol{x}\right]_1 \right\},\ldots,    \mathrm{Re}\left\{ \left[\boldsymbol{x}\right]_M \right\}  \right]^T$.

%% file: 02_System_Model.tex
\section{System Model}
\label{sec:system_model}

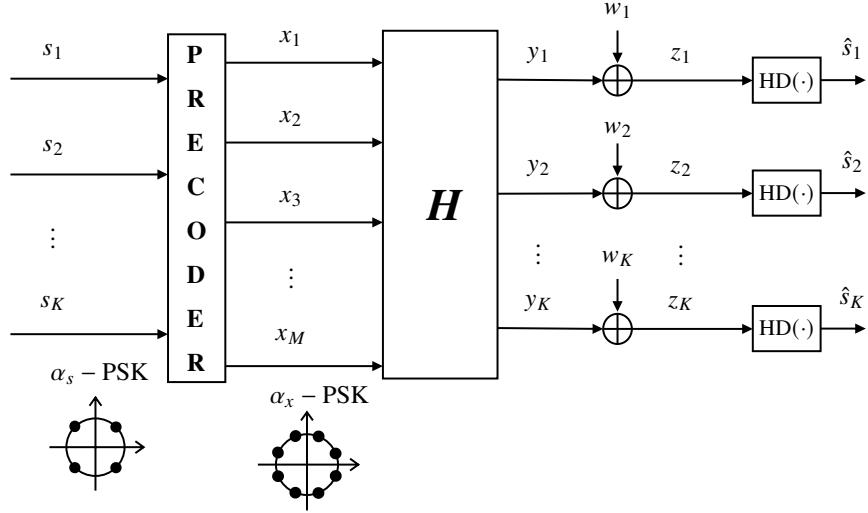
\begin{figure*} 
\centering
\input{figures/System_Model}
\caption{Multiuser MIMO downlink with discrete precoding}
\label{fig:system_model}       
\vspace{-1em}
\end{figure*}

The system model, illustrated in Fig.~\ref{fig:system_model}, consists in a single-cell MU-MIMO scenario where the BS has perfect channel state information (CSI) and is equipped with $M$ transmit antennas which serves $K$ single antenna users.

In this study, a symbol level transmission is considered where $s_k$ represents the data symbol to be delivered for the $k$-th user. Each symbol $s_k$ is considered to belong to the set  $\mathcal{S}$ that represents all possible symbols of a $\alpha_{s}$-PSK modulation and is given by 
\begin{align}
	    \mathcal{S}= \left\{s: s= e^  \frac{j\pi (2 i+1) }{\alpha_{s}}  \textrm{,  for  }  i=1,\ldots, \alpha_{s} \right\}  \textrm{.}
	    \label{S_set}
\end{align}
The symbols of all users are described in a stacked vector notation as $\boldsymbol{s}=[{s}_1,\ldots,{s}_K]^T \in \mathcal{S}^K$. It is considered that different users' symbols are independent meaning, $\text{P}\pc{s_i|s_j}=\text{P}\pc{s_i},\forall{i\neq j}$. Moreover it is assumed that all symbols have the same probability meaning $\text{P}\pc{s_k=s_i}=\frac{1}{\alpha_s},\forall i \in {1, \hdots, \alpha_s}$.
Based on $\boldsymbol{s}$ the precoder computes the transmit vector $\boldsymbol{x}=[x_{1} ,\hdots, x_{M}]^{T}$, whose entries are constrained to the set $\mathcal{X}$, given by
\begin{align}
	    \mathcal{X}= \left\{x: x= e^  \frac{j\pi (2 i+1) }{\alpha_{x}}  \textrm{,  for  }  i=1,\ldots, \alpha_{x} \right\} \textrm{,}
	    \label{X_set}
\end{align}
which describes an $\alpha_{x}$-PSK alphabet.
During this study the transmit vector $\boldsymbol{x}$ will, many times, be described using real-valued notation as follows
\begin{align}
\label{eq:rv_x}
    \boldsymbol{x}_{\text{r}}=\pr{ \mathrm{Re} \left\{\boldsymbol{x}_1\right\} \   \mathrm{Im} \left\{\boldsymbol{x}_1\right\} \
\cdots \
\mathrm{Re} \left\{\boldsymbol{x}_M\right\} \  \mathrm{Im} \left\{\boldsymbol{x}_M\right\}}^T.
\end{align}
The conversion operation from complex-valued to real-valued description is represented using the operator $R(\cdot)$. The opposite conversion, i.e. real-valued to complex-valued description, is denoted by the operator $C(\cdot)$.


The vector $\boldsymbol{x}$ is transmitted over a frequency flat fading channel described by the matrix $\boldsymbol{H}$ with coefficients $h_{k,m}=g_{k,m}\sqrt{\beta_k}$, where $g_{k,m}$ represents the complex small-scale fading between the $m$-th antenna and the $k$-th user, and ${\beta_k}$ denotes the real valued large-scale fading coefficient of the $k$-th user, $k=1...K$ and $m=1...M$.
With this, the received signal corresponding to the $k$-th user is given by
\begin{align}
\label{eq:symbols_at_receiver}
    {z_k} &= y_k+{w_k} \notag\\
    &= \boldsymbol{h}_k\ \boldsymbol{x}+{w_k}\text{,}
\end{align}
where $y_k$ is the noiseless received signal from the $k$-th user, $\boldsymbol{h}_k$ is the $k$-th row of the channel matrix $\boldsymbol{H}$ and the complex random variable ${w_k}\sim \mathcal{CN}  ({0},\sigma_w^2)$ represents additive white Gaussian noise (AWGN).
Using stacked vector notation equation \eqref{eq:symbols_at_receiver} can be extended to 
\begin{align}
    \boldsymbol{z} &= \boldsymbol{y}+\boldsymbol{w} \notag\\
    &=\boldsymbol{H}\ \boldsymbol{x}+\boldsymbol{w}\text{,}
\end{align}
where $\boldsymbol{z}=\pr{z_{1} \hdots z_{K}}^{T}$, $\boldsymbol{y}=\pr{y_{1} \hdots y_{K}}^{T}$ and $\boldsymbol{w}=\pr{w_{1} \hdots w_{K}}^{T}$.
Each received symbol $z_k$ is, then, hard detected based on which decision region it belongs, meaning that $z_k$ is detected as $s_i$ if $z_k \in \mathcal{S}_i$. In the case of PSK modulation the decision regions are circle sectors with infinite radius and angle of $2\theta$, where $\theta$ is given by $\theta=\frac{\pi}{\alpha_s}$. As such the estimated symbol from the $k$-th user is given by $\hat{s}_k=\text{HD}(z_k)$, where $\text{HD}(\cdot)$ represents the hard detection operation.
Finally, the estimated symbol vector can be written in stacked notation as $\hat{\boldsymbol{s}}=\pr{\hat{s}_1, \hdots, \hat{s}_K}$.

%% file: figures/System_Model.tex
\tikzset{every picture/.style={line width=0.75pt}} 

\begin{tikzpicture}[x=0.45pt,y=0.45pt,yscale=-1,xscale=1]

\draw   (306.46,75.44) -- (354.63,75.44) -- (354.63,368) -- (306.46,368) -- cycle ;
\draw    (174,112.76) -- (303.46,112.76) ;
\draw [shift={(306.46,112.76)}, rotate = 180] [fill={rgb, 255:red, 0; green, 0; blue, 0 }  ][line width=0.08]  [draw opacity=0] (8.93,-4.29) -- (0,0) -- (8.93,4.29) -- cycle    ;
\draw    (174,193.36) -- (303.46,193.36) ;
\draw [shift={(306.46,193.36)}, rotate = 180] [fill={rgb, 255:red, 0; green, 0; blue, 0 }  ][line width=0.08]  [draw opacity=0] (8.93,-4.29) -- (0,0) -- (8.93,4.29) -- cycle    ;
\draw    (174,327.7) -- (303.46,327.7) ;
\draw [shift={(306.46,327.7)}, rotate = 180] [fill={rgb, 255:red, 0; green, 0; blue, 0 }  ][line width=0.08]  [draw opacity=0] (8.93,-4.29) -- (0,0) -- (8.93,4.29) -- cycle    ;
\draw    (354.63,99.32) -- (484.09,99.32) ;
\draw [shift={(487.09,99.32)}, rotate = 180] [fill={rgb, 255:red, 0; green, 0; blue, 0 }  ][line width=0.08]  [draw opacity=0] (8.93,-4.29) -- (0,0) -- (8.93,4.29) -- cycle    ;
\draw    (354.63,166.49) -- (484.09,166.49) ;
\draw [shift={(487.09,166.49)}, rotate = 180] [fill={rgb, 255:red, 0; green, 0; blue, 0 }  ][line width=0.08]  [draw opacity=0] (8.93,-4.29) -- (0,0) -- (8.93,4.29) -- cycle    ;
\draw    (354.63,233.66) -- (484.09,233.66) ;
\draw [shift={(487.09,233.66)}, rotate = 180] [fill={rgb, 255:red, 0; green, 0; blue, 0 }  ][line width=0.08]  [draw opacity=0] (8.93,-4.29) -- (0,0) -- (8.93,4.29) -- cycle    ;
\draw    (354.63,354.57) -- (484.09,354.57) ;
\draw [shift={(487.09,354.57)}, rotate = 180] [fill={rgb, 255:red, 0; green, 0; blue, 0 }  ][line width=0.08]  [draw opacity=0] (8.93,-4.29) -- (0,0) -- (8.93,4.29) -- cycle    ;
\draw   (487.09,72.45) -- (583.42,72.45) -- (583.42,365.01) -- (487.09,365.01) -- cycle ;
\draw    (583.42,209.54) -- (669.25,209.28) ;
\draw [shift={(672.25,209.27)}, rotate = 539.8299999999999] [fill={rgb, 255:red, 0; green, 0; blue, 0 }  ][line width=0.08]  [draw opacity=0] (8.93,-4.29) -- (0,0) -- (8.93,4.29) -- cycle    ;
\draw   (672.25,209.27) .. controls (672.25,202.27) and (677.64,196.59) .. (684.29,196.59) .. controls (690.94,196.59) and (696.34,202.27) .. (696.34,209.27) .. controls (696.34,216.28) and (690.94,221.96) .. (684.29,221.96) .. controls (677.64,221.96) and (672.25,216.28) .. (672.25,209.27) -- cycle ; \draw   (672.25,209.27) -- (696.34,209.27) ; \draw   (684.29,196.59) -- (684.29,221.96) ;
\draw    (684.29,167.23) -- (684.29,191.84) ;
\draw [shift={(684.29,194.84)}, rotate = 270] [fill={rgb, 255:red, 0; green, 0; blue, 0 }  ][line width=0.08]  [draw opacity=0] (8.93,-4.29) -- (0,0) -- (8.93,4.29) -- cycle    ;
\draw    (696.34,209.27) -- (795,209.27) ;
\draw [shift={(798,209.27)}, rotate = 180] [fill={rgb, 255:red, 0; green, 0; blue, 0 }  ][line width=0.08]  [draw opacity=0] (8.93,-4.29) -- (0,0) -- (8.93,4.29) -- cycle    ;
\draw  (206.78,422.75) -- (286,422.75)(245.47,380.93) -- (245.47,459.29) (279,417.75) -- (286,422.75) -- (279,427.75) (240.47,387.93) -- (245.47,380.93) -- (250.47,387.93)  ;
\draw   (221.02,422.75) .. controls (221.02,409.4) and (231.97,398.57) .. (245.47,398.57) .. controls (258.98,398.57) and (269.92,409.4) .. (269.92,422.75) .. controls (269.92,436.11) and (258.98,446.94) .. (245.47,446.94) .. controls (231.97,446.94) and (221.02,436.11) .. (221.02,422.75) -- cycle ;
\draw  [color={rgb, 255:red, 0; green, 0; blue, 0 }  ,draw opacity=1 ][fill={rgb, 255:red, 0; green, 0; blue, 0 }  ,fill opacity=1 ] (259.94,403.06) .. controls (261.46,401.54) and (263.94,401.54) .. (265.47,403.06) .. controls (267,404.57) and (267,407.02) .. (265.47,408.53) .. controls (263.94,410.04) and (261.46,410.04) .. (259.94,408.53) .. controls (258.41,407.02) and (258.41,404.57) .. (259.94,403.06) -- cycle ;
\draw  [color={rgb, 255:red, 0; green, 0; blue, 0 }  ,draw opacity=1 ][fill={rgb, 255:red, 0; green, 0; blue, 0 }  ,fill opacity=1 ] (225.36,437.26) .. controls (226.88,435.75) and (229.36,435.75) .. (230.89,437.26) .. controls (232.42,438.77) and (232.42,441.22) .. (230.89,442.73) .. controls (229.36,444.24) and (226.88,444.24) .. (225.36,442.73) .. controls (223.83,441.22) and (223.83,438.77) .. (225.36,437.26) -- cycle ;
\draw  [color={rgb, 255:red, 0; green, 0; blue, 0 }  ,draw opacity=1 ][fill={rgb, 255:red, 0; green, 0; blue, 0 }  ,fill opacity=1 ] (225.36,402.51) .. controls (226.78,400.89) and (229.17,400.81) .. (230.7,402.32) .. controls (232.22,403.83) and (232.31,406.36) .. (230.89,407.98) .. controls (229.47,409.6) and (227.08,409.68) .. (225.55,408.17) .. controls (224.02,406.66) and (223.94,404.13) .. (225.36,402.51) -- cycle ;
\draw  [color={rgb, 255:red, 0; green, 0; blue, 0 }  ,draw opacity=1 ][fill={rgb, 255:red, 0; green, 0; blue, 0 }  ,fill opacity=1 ] (259.94,437.26) .. controls (261.46,435.75) and (263.94,435.75) .. (265.47,437.26) .. controls (267,438.77) and (267,441.22) .. (265.47,442.73) .. controls (263.94,444.24) and (261.46,444.24) .. (259.94,442.73) .. controls (258.41,441.22) and (258.41,438.77) .. (259.94,437.26) -- cycle ;

\draw  (382,437.42) -- (466.47,437.42)(423.26,393.26) -- (423.26,476) (459.47,432.42) -- (466.47,437.42) -- (459.47,442.42) (418.26,400.26) -- (423.26,393.26) -- (428.26,400.26)  ;
\draw   (397.19,437.42) .. controls (397.19,423.31) and (408.86,411.88) .. (423.26,411.88) .. controls (437.66,411.88) and (449.33,423.31) .. (449.33,437.42) .. controls (449.33,451.52) and (437.66,462.96) .. (423.26,462.96) .. controls (408.86,462.96) and (397.19,451.52) .. (397.19,437.42) -- cycle ;
\draw  [color={rgb, 255:red, 0; green, 0; blue, 0 }  ,draw opacity=1 ][fill={rgb, 255:red, 0; green, 0; blue, 0 }  ,fill opacity=1 ] (445.74,424.26) .. controls (447.87,423.41) and (450.3,424.42) .. (451.16,426.52) .. controls (452.03,428.61) and (450.99,430.99) .. (448.86,431.83) .. controls (446.72,432.68) and (444.29,431.67) .. (443.43,429.58) .. controls (442.57,427.48) and (443.6,425.1) .. (445.74,424.26) -- cycle ;
\draw  [color={rgb, 255:red, 0; green, 0; blue, 0 }  ,draw opacity=1 ][fill={rgb, 255:red, 0; green, 0; blue, 0 }  ,fill opacity=1 ] (397.39,443.39) .. controls (399.53,442.55) and (401.96,443.56) .. (402.82,445.65) .. controls (403.68,447.74) and (402.65,450.12) .. (400.51,450.97) .. controls (398.38,451.81) and (395.95,450.8) .. (395.09,448.71) .. controls (394.22,446.62) and (395.25,444.24) .. (397.39,443.39) -- cycle ;
\draw  [color={rgb, 255:red, 0; green, 0; blue, 0 }  ,draw opacity=1 ][fill={rgb, 255:red, 0; green, 0; blue, 0 }  ,fill opacity=1 ] (412.03,409.61) .. controls (414.1,408.62) and (416.48,409.51) .. (417.35,411.61) .. controls (418.21,413.7) and (417.23,416.2) .. (415.15,417.19) .. controls (413.08,418.18) and (410.69,417.29) .. (409.83,415.2) .. controls (408.97,413.11) and (409.95,410.6) .. (412.03,409.61) -- cycle ;
\draw  [color={rgb, 255:red, 0; green, 0; blue, 0 }  ,draw opacity=1 ][fill={rgb, 255:red, 0; green, 0; blue, 0 }  ,fill opacity=1 ] (431.33,457.5) .. controls (433.46,456.66) and (435.9,457.67) .. (436.76,459.76) .. controls (437.62,461.85) and (436.59,464.23) .. (434.45,465.08) .. controls (432.32,465.93) and (429.89,464.91) .. (429.02,462.82) .. controls (428.16,460.73) and (429.19,458.35) .. (431.33,457.5) -- cycle ;

\draw  [color={rgb, 255:red, 0; green, 0; blue, 0 }  ,draw opacity=1 ][fill={rgb, 255:red, 0; green, 0; blue, 0 }  ,fill opacity=1 ] (429.16,412.16) .. controls (430.03,410.07) and (432.46,409.05) .. (434.59,409.9) .. controls (436.73,410.74) and (437.76,413.13) .. (436.9,415.22) .. controls (436.04,417.31) and (433.6,418.32) .. (431.47,417.48) .. controls (429.33,416.63) and (428.3,414.25) .. (429.16,412.16) -- cycle ;
\draw  [color={rgb, 255:red, 0; green, 0; blue, 0 }  ,draw opacity=1 ][fill={rgb, 255:red, 0; green, 0; blue, 0 }  ,fill opacity=1 ] (409.63,459.52) .. controls (410.49,457.42) and (412.92,456.41) .. (415.06,457.26) .. controls (417.2,458.1) and (418.23,460.48) .. (417.36,462.58) .. controls (416.5,464.67) and (414.07,465.68) .. (411.94,464.83) .. controls (409.8,463.99) and (408.77,461.61) .. (409.63,459.52) -- cycle ;
\draw  [color={rgb, 255:red, 0; green, 0; blue, 0 }  ,draw opacity=1 ][fill={rgb, 255:red, 0; green, 0; blue, 0 }  ,fill opacity=1 ] (394.99,425.74) .. controls (395.71,423.59) and (398.02,422.53) .. (400.15,423.37) .. controls (402.29,424.22) and (403.44,426.65) .. (402.73,428.8) .. controls (402.02,430.95) and (399.71,432.01) .. (397.57,431.16) .. controls (395.43,430.32) and (394.28,427.89) .. (394.99,425.74) -- cycle ;
\draw  [color={rgb, 255:red, 0; green, 0; blue, 0 }  ,draw opacity=1 ][fill={rgb, 255:red, 0; green, 0; blue, 0 }  ,fill opacity=1 ] (443.57,445.4) .. controls (444.43,443.31) and (446.86,442.3) .. (449,443.14) .. controls (451.13,443.99) and (452.17,446.37) .. (451.3,448.46) .. controls (450.44,450.56) and (448.01,451.57) .. (445.88,450.72) .. controls (443.74,449.88) and (442.71,447.5) .. (443.57,445.4) -- cycle ;

\draw   (798,190) -- (855,190) -- (855,228) -- (798,228) -- cycle ;
\draw    (583.42,113.54) -- (669.25,114.25) ;
\draw [shift={(672.25,114.27)}, rotate = 180.47] [fill={rgb, 255:red, 0; green, 0; blue, 0 }  ][line width=0.08]  [draw opacity=0] (8.93,-4.29) -- (0,0) -- (8.93,4.29) -- cycle    ;
\draw   (672.25,114.27) .. controls (672.25,107.27) and (677.64,101.59) .. (684.29,101.59) .. controls (690.94,101.59) and (696.34,107.27) .. (696.34,114.27) .. controls (696.34,121.28) and (690.94,126.96) .. (684.29,126.96) .. controls (677.64,126.96) and (672.25,121.28) .. (672.25,114.27) -- cycle ; \draw   (672.25,114.27) -- (696.34,114.27) ; \draw   (684.29,101.59) -- (684.29,126.96) ;
\draw    (684.29,72.23) -- (684.29,96.84) ;
\draw [shift={(684.29,99.84)}, rotate = 270] [fill={rgb, 255:red, 0; green, 0; blue, 0 }  ][line width=0.08]  [draw opacity=0] (8.93,-4.29) -- (0,0) -- (8.93,4.29) -- cycle    ;
\draw    (696.34,114.27) -- (795,114.27) ;
\draw [shift={(798,114.27)}, rotate = 180] [fill={rgb, 255:red, 0; green, 0; blue, 0 }  ][line width=0.08]  [draw opacity=0] (8.93,-4.29) -- (0,0) -- (8.93,4.29) -- cycle    ;
\draw   (798,95) -- (855,95) -- (855,133) -- (798,133) -- cycle ;
\draw    (583.42,322.54) -- (669.25,323.25) ;
\draw [shift={(672.25,323.27)}, rotate = 180.47] [fill={rgb, 255:red, 0; green, 0; blue, 0 }  ][line width=0.08]  [draw opacity=0] (8.93,-4.29) -- (0,0) -- (8.93,4.29) -- cycle    ;
\draw   (672.25,323.27) .. controls (672.25,316.27) and (677.64,310.59) .. (684.29,310.59) .. controls (690.94,310.59) and (696.34,316.27) .. (696.34,323.27) .. controls (696.34,330.28) and (690.94,335.96) .. (684.29,335.96) .. controls (677.64,335.96) and (672.25,330.28) .. (672.25,323.27) -- cycle ; \draw   (672.25,323.27) -- (696.34,323.27) ; \draw   (684.29,310.59) -- (684.29,335.96) ;
\draw    (684.29,281.23) -- (684.29,305.84) ;
\draw [shift={(684.29,308.84)}, rotate = 270] [fill={rgb, 255:red, 0; green, 0; blue, 0 }  ][line width=0.08]  [draw opacity=0] (8.93,-4.29) -- (0,0) -- (8.93,4.29) -- cycle    ;
\draw    (696.34,323.27) -- (795,323.27) ;
\draw [shift={(798,323.27)}, rotate = 180] [fill={rgb, 255:red, 0; green, 0; blue, 0 }  ][line width=0.08]  [draw opacity=0] (8.93,-4.29) -- (0,0) -- (8.93,4.29) -- cycle    ;
\draw   (798,304) -- (855,304) -- (855,342) -- (798,342) -- cycle ;
\draw    (855,114) -- (888,114) -- (890,114) ;
\draw [shift={(893,114)}, rotate = 180] [fill={rgb, 255:red, 0; green, 0; blue, 0 }  ][line width=0.08]  [draw opacity=0] (8.93,-4.29) -- (0,0) -- (8.93,4.29) -- cycle    ;
\draw    (855,209) -- (888,209) -- (890,209) ;
\draw [shift={(893,209)}, rotate = 180] [fill={rgb, 255:red, 0; green, 0; blue, 0 }  ][line width=0.08]  [draw opacity=0] (8.93,-4.29) -- (0,0) -- (8.93,4.29) -- cycle    ;
\draw    (855,323) -- (888,323) -- (890,323) ;
\draw [shift={(893,323)}, rotate = 180] [fill={rgb, 255:red, 0; green, 0; blue, 0 }  ][line width=0.08]  [draw opacity=0] (8.93,-4.29) -- (0,0) -- (8.93,4.29) -- cycle    ;

\draw (209.7,241.7) node  [scale=0.8] {$\vdots $};
\draw (410,275.13) node  [scale=0.8] {$\vdots $};
\draw (617.77,255.27) node  [scale=0.8] {$\vdots $};
\draw (737.61,256.04) node  [scale=0.8] {$\vdots $};
\draw (209.7,90.19) node  [scale=0.8] {$s_{1}$};
\draw (209.7,170.8) node  [scale=0.8] {$s_{2}$};
\draw (209.7,301.46) node  [scale=0.8] {$s_{K}$};
\draw (410,79.25) node  [scale=0.8] {$x_{1}$};
\draw (410,146.42) node  [scale=0.8] {$x_{2}$};
\draw (410,213.59) node  [scale=0.8] {$x_{3}$};
\draw (410,329.72) node  [scale=0.8] {$x_{M}$};
\draw (617.77,188.05) node  [scale=0.8] {$y_{2}$};
\draw (737.61,188.05) node  [scale=0.8] {$z_{2}$};
\draw (684.91,155.26) node  [scale=0.8]  {$w_{2}$};
\draw (330.54,221.72) node  [scale=0.8] [align=left] {\textbf{P}\\\textbf{R}\\\textbf{E}\\\textbf{C}\\\textbf{O}\\\textbf{D}\\\textbf{E}\\\textbf{R}};
\draw (539.51,218.56) node  [font=\Large] [align=left] {\textit{\textbf{H}}};
\draw (249.09,361) node  [scale=0.8]  {$\alpha _{s} - \text{PSK}$};
\draw (433.82,380.06) node  [scale=0.8]  {$\alpha _{x} -\text{PSK}$};
\draw (617.77,93.05) node  [scale=0.8] {$y_{1}$};
\draw (737.61,93.05) node  [scale=0.8] {$z_{1}$};
\draw (684.91,55.26) node  [scale=0.8]  {$w_{1}$};
\draw (617.77,301.46) node  [scale=0.8] {$y_{K}$};
\draw (737.61,301.46) node  [scale=0.8] {$z_{K}$};
\draw (684.91,264.26) node  [scale=0.8]  {$w_{K}$};
\draw (827,115.5) node  [scale=0.7]  {$\text{HD}( \cdot )$};
\draw (827,210) node  [scale=0.7]  {$\text{HD}( \cdot )$};
\draw (827,324) node  [scale=0.7]  {$\text{HD}( \cdot )$};
\draw (881.5,87.5) node  [scale=0.8] {$\hat{s}_{1}$};
\draw (881.5,182.5) node  [scale=0.8] {$\hat{s}_{2}$};
\draw (881.5,296.5) node  [scale=0.8] {$\hat{s}_{K}$};

\end{tikzpicture}

%% file: Formulations/Formulations.tex
\section{MSEP Precoding Formulation}
\label{sec:precoding_formulation}

In this study we consider as the precoding design criterion the minimization of the SEP, or, equivalently, the maximization of the probability of correct detection. 
The probability of detecting the data vector $\boldsymbol{s}$ conditioned on the transmit vector $\boldsymbol{x}$ can be computed based on the probabilities of detection of the individual users as
\begin{align}
\label{eq:probability_definition1}
    \text{P}(\hat{\boldsymbol{s}}=\boldsymbol{s}|\boldsymbol{x})&= \displaystyle \prod_{k=1}^K \text{P}(\hat{s}_k=s_k|\boldsymbol{x})\ .
\end{align}
To simplify the notation we denote $\text{P}(\hat{\boldsymbol{s}}=\boldsymbol{s}|\boldsymbol{x})$ as $\text{P}(\hat{\boldsymbol{s}}|\boldsymbol{x})$ and $\text{P}(\hat{s}_k=s_k|\boldsymbol{x})$ as $\text{P}(\hat{s}_k|\boldsymbol{x})$. With this, equation \eqref{eq:probability_definition1} is rewritten as
\begin{align}
\label{eq:probability_definition}
    \text{P}(\hat{\boldsymbol{s}}|\boldsymbol{x})&= \displaystyle \prod_{k=1}^K \text{P}(\hat{s}_k|\boldsymbol{x})\ .
\end{align}
As stated before, the detector decides for $s_k$ when the received symbol $z_k$ belongs to $\mathcal{S}_k$. Thus, the individual user probabilities are given by
\begin{align}
\label{eq:prob_gaussian}
    \text{P}\pc{\hat{s}_k|\boldsymbol{x}}&=\text{P}\pc{z_k \in \mathcal{S}_k|\boldsymbol{x}}=\frac{1}{\pi \sigma_w^2} \int \int_{\mathcal{S}_k}  \text{e}^{-\frac{\PM{\PM{\boldsymbol{t}-{y}_{k}}}_2^2}{\sigma_w^2}} d\boldsymbol{t}.
\end{align}
The integral from \eqref{eq:prob_gaussian}, leads to two different formulations for the correct detection probability. The first corresponds to the specific case of QPSK the data modulation ($\alpha_s=4$) and the second applies for any value of $\alpha_s$. 

\input{Formulations/QMSEP_formulation}

\input{Formulations/UB-MSEP_formulation}

\subsection{Considerations about the convexity of the MSEP formulations}

Note that, the QMSEP objective function is convex, since $-\log(\Phi(\boldsymbol{Ax}))$ is convex in $\boldsymbol{x}$ and the sum of convex function is also convex, c.f. \cite{Boyd_2004}. 
As shown in appendix~\ref{app:Ub_msep} the objective function of the UBMSEP formulation is convex for $(d_{1,k}, d_{2,k}) >(0,0),  k\in{1,\hdots,K}$. To ensure a convex objective, the original UBMSEP problem is rewritten as 
\begin{align}
\label{eq:reformulated_rv_ubmsep_prob}
    &\displaystyle \min_{\boldsymbol{x}_\text{r}} \  -\sum_{k=1}^K \log \pc{ \text{erf}\pc{ \pc{\boldsymbol{h}_{\text{R},\theta,k}^{s^*} -\boldsymbol{h}_{\text{I},\theta,k}^{s^*}} \boldsymbol{x}_{\text{r}} }+
    \text{erf}\pc{ \pc{\boldsymbol{h}_{\text{R},\theta,k}^{s^*} +\boldsymbol{h}_{\text{I},\theta,k}^{s^*}} \boldsymbol{x}_{\text{r} } }} \\
   &\text{s.t.} \ \ \ \  \begin{bmatrix}  \boldsymbol{H}_{\text{R},\theta}^{s^*} - \boldsymbol{H}_{\text{I},\theta}^{s^*} \\ \boldsymbol{H}_{\text{R},\theta}^{s^*} + \boldsymbol{H}_{\text{I},\theta}^{s^*} 
\end{bmatrix} \boldsymbol{x}_\text{r} \geq \boldsymbol{0},   \ \ \ \ {x}_{\text{r},2m-1}+j{x}_{\text{r},2m} \in \mathcal{X} \ \ \textrm{for } m=1,\ldots,M\textrm{.} 
\notag
\end{align}
Note that, the optimal solution from \eqref{eq:msep_prob_psk} is not necessarily the same as the optimal from \eqref{eq:reformulated_rv_ubmsep_prob}. However, different solutions are only possible if, for at least one user, the optimal of \eqref{eq:msep_prob_psk} yields a noiseless received symbol ${y}_i$ in the incorrect decision region. This leads to an optimal SEP, for this user, greater than half. 
This is, in general, not an interesting case, since future wireless communications systems will be designed to provide high reliability and, with this, avoid this kind of scenario. 

Both MSEP optimization problems describe the minimization of a convex objective under a discrete feasible set. Note that, due to the discrete constraint $x_{\text{r},2m}+j{x}_{\text{r},2m+1} \in \mathcal{X} \ \ \textrm{for } m=1,\ldots,M$ the MSEP problems are classified as a discrete programming problem (DPP). In the following sections optimal and suboptimal methods to solve the MSEP DPPs are proposed.

%% file: Formulations/QMSEP_formulation.tex
\subsection{QMSEP Problem Formulation}

In this subsection, we expose the QPSK MSEP (QMSEP) formulation, which was first presented in \cite{mezghani2020massive}. When the data modulation is QPSK, $\text{Re}\chav{s_k}$ and $\text{Im}\chav{s_k}$ are independent and, thus, the decision regions $\mathcal{S}_k$ can be written as $\mathcal{R}_k \cap \mathcal{I}_k$, where $\mathcal{R}_k$ and $\mathcal{I}_k$ are the decision regions the real and imaginary parts of $s_k$. The probability of the detector deciding for $s_k$ is given by
\begin{align}
\label{eq:prob_qmsep}
    \text{P}\pc{\hat{s}_k|\boldsymbol{x}}&=\text{P}\pc{z_k \in \mathcal{S}_k|\boldsymbol{x}} \notag\\
    \text{P}\pc{\hat{s}_k|\boldsymbol{x}}&=\text{P}\pc{\text{Re}\chav{z_k} \in \mathcal{R}_k|\boldsymbol{x}}\text{P}\pc{\text{Im}\chav{z_k} \in \mathcal{I}_k|\boldsymbol{x}}
\end{align}
 With this, for $s_k=e^{j\frac{\pi}{4}}$,
\begin{align}
\label{eq:1bit_user_probability}
    \text{P}\pc{z_k \in \mathcal{R}|\boldsymbol{x}}&=\int_{0}^\infty \frac{1}{\sqrt{\pi \sigma_w^2}}\  e^{\frac{\pc{t-\text{Re}\chav{\boldsymbol{h}_k \boldsymbol{x}}}^2}{\sigma_w^2}} dt=\Phi\pc{\frac{\sqrt{2}\ \text{Re}\chav{\boldsymbol{h}_k \boldsymbol{x}} }{\sigma_w}},\notag\\
    \text{P}\pc{z_k \in \mathcal{I}|\boldsymbol{x}}&=\int_{0}^\infty \frac{1}{\sqrt{\pi \sigma_w^2}}\  e^{\frac{\pc{t-\text{Im}\chav{\boldsymbol{h}_k \boldsymbol{x}}}^2}{\sigma_w^2}} dt=\Phi\pc{\frac{\sqrt{2}\ \text{Im}\chav{\boldsymbol{h}_k \boldsymbol{x}} }{\sigma_w}}\notag.
\end{align}
An expression for the correct decision probability can be achieved for all elements in $\mathcal{S}$ as
\begin{align}
    \text{P}\pc{\hat{s}_k|\boldsymbol{x}}=\Phi\pc{\frac{\sqrt{2}\ \text{sign}\pc{\text{Re}\chav{s_k}} \text{Re}\chav{\boldsymbol{h}_k \boldsymbol{x}} }{\sigma_w}} \Phi\pc{\frac{\sqrt{2}\ \text{sign}\pc{\text{Im}\chav{s_k}} \text{Im}\chav{\boldsymbol{h}_k \boldsymbol{x}} }{\sigma_w}}.
\end{align}
With this, $\text{P}\pc{\boldsymbol{s}|\boldsymbol{x}}$ is computed, for the QPSK case, by inserting \eqref{eq:1bit_user_probability} into \eqref{eq:probability_definition}, which reads as 
\begin{align}
\label{eq:1bit_probability}
    \text{P}\pc{\hat{\boldsymbol{s}}|\boldsymbol{x}}=\displaystyle\prod_{k=1}^K\Phi\pc{\frac{\sqrt{2}\ \text{sign}\pc{\text{Re}\chav{s_k}} \text{Re}\chav{\boldsymbol{h}_k \boldsymbol{x}} }{\sigma_w}} \Phi\pc{\frac{\sqrt{2}\ \text{sign}\pc{\text{Im}\chav{s_k}} \text{Im}\chav{\boldsymbol{h}_k \boldsymbol{x}} }{\sigma_w}}.
\end{align}
Based on \eqref{eq:1bit_probability} an optimization problem can be written for maximizing $\text{P}\pc{\boldsymbol{s}|\boldsymbol{x}}$, which is cast as 
\begin{align}
\label{eq:1bit_opt_problem_intermediate}
    \arg \max_{\boldsymbol{x}\in \mathcal{X}^M}\displaystyle\prod_{k=1}^K \Phi\pc{\frac{\sqrt{2}\ \text{sign}\pc{\text{Re}\chav{s_k}} \text{Re}\chav{\boldsymbol{h}_k \boldsymbol{x}} }{\sigma_w}} \Phi\pc{\frac{\sqrt{2}\ \text{sign}\pc{\text{Im}\chav{s_k}} \text{Im}\chav{\boldsymbol{h}_k \boldsymbol{x}} }{\sigma_w}}.
\end{align}
Since the $\log(\cdot)$ is a monotonically increasing function, applying it to the objective from \eqref{eq:1bit_opt_problem_intermediate} does not change the optimal solution. The QPSK MSEP (QMSEP) optimization problem is, then, given by
\begin{align}
\label{eq:1bit_opt_problem}
    \arg \min_{\boldsymbol{x}\in \mathcal{X}^M}-\displaystyle\sum_{k=1}^K \log \pc{\Phi\pc{\frac{\sqrt{2}\ \text{sign}\pc{\text{Re}\chav{s_k}} \text{Re}\chav{\boldsymbol{h}_k \boldsymbol{x}} }{\sigma_w}}}+ \log\pc{\Phi\pc{\frac{\sqrt{2}\ \text{sign}\pc{\text{Im}\chav{s_k}} \text{Im}\chav{\boldsymbol{h}_k \boldsymbol{x}} }{\sigma_w}}}.
\end{align}
An alternative real valued formulation of the QMSEP optimization problem can be cast as  
\begin{align}
\label{eq:rv_qmsep_prob}
    &\min_{\boldsymbol{x}_\text{r}} \  -\sum_{k=1}^K \log \pc{\Phi\pc{\boldsymbol{h}_{\text{R},k} \boldsymbol{x}_\text{r}} }+ \log\pc{\Phi\pc{\boldsymbol{h}_{\text{I},k} \boldsymbol{x}_\text{r}}} \\
    &\text{s.t.} \ \ \ \  {x}_{\text{r},2m-1}+j{x}_{\text{r},2m} \in \mathcal{X} \ \ \ \ \textrm{for } m=1,\ldots,M     \textrm{.} \notag
\end{align}
where $\boldsymbol{h}_{\text{R},k}$ and $\boldsymbol{h}_{\text{I},k}$ are the $k$-th rows of matrices $\boldsymbol{H}_\text{R}$ and $\boldsymbol{H}_\text{I}$, respectively, which are defined as
$\boldsymbol{H}_\text{R}=\frac{\sqrt{2}}{\sigma_w} \text{diag}(\text{sign}(\text{Re}\chav{\boldsymbol{s}})) \boldsymbol{H}_\text{R}^Q$ and $\boldsymbol{H}_\text{I}=\frac{\sqrt{2}}{\sigma_w} \text{diag}(\text{sign}(\text{Im}\chav{\boldsymbol{s}})) \boldsymbol{H}_\text{I}^Q$, with
\begin{align}
&\boldsymbol{H}_{\text{R}}^Q    =      \begin{bmatrix} 
&\textrm{Re}\chav{{h}_{11}} &-\textrm{Im}\chav{{h}_{11}} 
                         &\cdots
&\textrm{Re}\chav{{h}_{1M}} \ &-\textrm{Im}\chav{{h}_{1M}}  \\
&\vdots &\vdots&\ddots&\vdots &\vdots&\\
&\textrm{Re}\chav{{h}_{K1}} &-\textrm{Im}\chav{{h}_{K1}} 
                         &\cdots
&\textrm{Re}\chav{{h}_{KM}} \ &-\textrm{Im}\chav{{h}_{KM}}  
\end{bmatrix},\\
&\boldsymbol{H}_{\text{I}}^Q    =      \begin{bmatrix} 
&\textrm{Im}\chav{{h}_{11}} &\textrm{Re}\chav{{h}_{11}} 
                         &\cdots
&\textrm{Im}\chav{{h}_{1M}} \ &\textrm{Re}\chav{{h}_{1M}}  \\
&\vdots &\vdots&\ddots&\vdots &\vdots&\\
&\textrm{Im}\chav{{h}_{K1}} &\textrm{Re}\chav{{h}_{K1}} 
                         &\cdots
&\textrm{Re}\chav{{h}_{KM}} \ &\textrm{Im}\chav{{h}_{KM}} 
\end{bmatrix}.
\end{align}
The QMSEP formulation is limited for QPSK data modulation ($\alpha_s=4$). In the following, we devise the MSEP formulation for higher order PSK modulations.

%% file: Formulations/UB-MSEP_formulation.tex
\subsection{MSEP Problem formulation based on the Union Bound probability}

\begin{figure}[h] 
\centering
\input{figures/Union_bound}
\caption{Representation of the Union Bound}
\label{fig:union_bound}       
\end{figure}
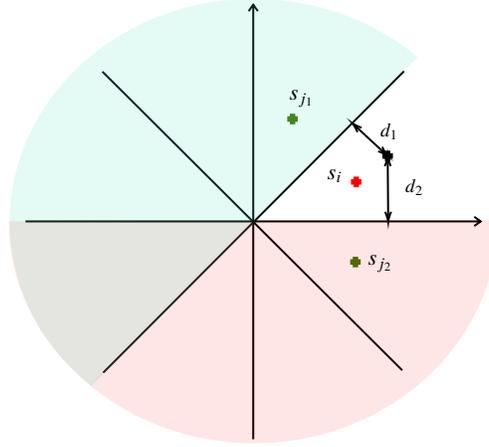

When higher order PSK modulations are used, the decision regions cannot be written as the intersection of two independent regions. As such, the exact computation of the integral presented in \eqref{eq:prob_gaussian} would lead to precoding algorithms with prohibitive computational complexity. 
Thus, for the MSEP formulation with $\alpha_s \neq 4$, we consider the maximization of a lower bound on the correct detection probability.
The Union Bound, also known Boole's inequality \cite{Boole}, states that for any finite or countable set of events, the probability that at least one of the events happens is smaller or equal than the sum of the probabilities of the individual events, meaning
\begin{align}
    {\displaystyle \text{P} \left(\bigcup _{i}A_{i}\right)\leq \sum _{i}{\text{P} }(A_{i}).}
\end{align}
With this, the error probability for the $k$-th user, $\text{P}_\text{e}\pc{\hat{s}_k|\boldsymbol{x}}$, can be bounded by 
\begin{align}
\label{eq:union_bound}
        \text{P}_\text{e}\pc{\hat{s}_k|\boldsymbol{x}}& =\text{P}\pc{z_k \in \mathcal{Z}_1 \cup\mathcal{Z}_2 |\boldsymbol{x}} \notag\\
        &\leq \text{P}\pc{z_k \in \mathcal{Z}_1|\boldsymbol{x}} + \text{P}\pc{z_k \in \mathcal{Z}_2|\boldsymbol{x}},
\end{align}
where the sets $\mathcal{Z}_1$ and $\mathcal{Z}_2$ are depicted in Fig.~\ref{fig:union_bound}.
When using the union bound, the error probability can be easily computed based on the minimum distances to the decision thresholds (MDDTs), $d_{1,k}$ and $d_{2,k}$, as  
\begin{align}
    \text{P}\pc{z_k \in \mathcal{Z}_1|\boldsymbol{x}}&=\int_{d_{1,k}}^\infty \frac{1}{\sqrt{\pi \sigma_w^2}}\  e^{\frac{t^2}{\sigma_w^2}} dt=\text{Q}\pc{\frac{\sqrt{2}\  d_{1,k}}{\sigma_w}}\\
    \text{P}\pc{z_k \in \mathcal{Z}_2|\boldsymbol{x}}&=\int_{d_{2,k}}^\infty \frac{1}{\sqrt{\pi \sigma_w^2}}\  e^{\frac{t^2}{\sigma_w^2}} dt
    =\text{Q}\pc{\frac{\sqrt{2}\  d_{2,k}}{\sigma_w}}.
\end{align}
The MDDTs are computed, similarly to in \cite{MSM_precoder} and \cite{General_MMDDT_BB}, by applying a rotation of $\text{arg}\{s_k^*\}=-\phi_{s_k}$ to the coordinate system such that the symbol of interest is placed on the real axis. 
This is done by multiplying both $s_k$ and $y_k$ by $e^{-j\phi_{s_k}}$ which results in
\begin{align}
    \quad e^{-j\phi_{s_k}} s_k=1, \quad \omega_k= e^{-j\phi_{s_k}} y_k .
\end{align}
Based on the rotated coordinate system the MDDTs are computed as
\begin{align}
    d_{1,k}&=\text{Re}\chav{\omega_k} \sin{\theta}-\PM{\text{Im}\chav{\omega_k}} \cos{\theta}\\
    d_{2,k}&=\text{Re}\chav{\omega_k} \sin{\theta}+\PM{\text{Im}\chav{\omega_k}} \cos{\theta},
\end{align}
which leads to an equivalent formulation as defining 
\begin{align}
    d_{1,k}&=\text{Re}\chav{s_k^* \boldsymbol{h}_k\boldsymbol{x}} \sin{\theta}-\text{Im}\chav{s_k^* \boldsymbol{h}_k\boldsymbol{x}} \cos{\theta}\\
    d_{2,k}&=\text{Re}\chav{s_k^* \boldsymbol{h}_k\boldsymbol{x}} \sin{\theta}+\text{Im}\chav{s_k^* \boldsymbol{h}_k\boldsymbol{x}} \cos{\theta}.
\end{align}
Based on the $d_{1,k}$ and $d_{2,k}$ one can construct a bound on the correct detection probability of the $k$-th user as
\begin{align}
\label{eq:individual_error_probability}
    \text{P}\pc{\hat{s}_k|\boldsymbol{x}}& = 1-{\text{P}_\text{e}}\pc{\hat{s}_k|\boldsymbol{x}}\notag\\
    &\geq 1-\pc{\text{P}\pc{z_k \in \mathcal{Z}_1|\boldsymbol{x}} + \text{P}\pc{z_k \in \mathcal{Z}_2|\boldsymbol{x}}}\notag\\
    &= {1- \text{Q}\pc{\frac{\sqrt{2} \ d_{1,k}}{\sigma_w}} - \text{Q}\pc{\frac{\sqrt{2}\ d_{2,k}}{\sigma_w}}} \\ \notag
    &= \frac{1}{2} \text{erf} \pc{ \frac{d_{1,k}}{\sigma_w}} + \frac{1}{2} \text{erf} \pc{ \frac{d_{2,k}}{\sigma_w}}
\end{align}
 With this, a bound for $\text{P}(\boldsymbol{s}|\boldsymbol{x})$ is computed by inserting \eqref{eq:individual_error_probability} into \eqref{eq:probability_definition}, which reads as 
\begin{align}
\label{eq:Prob_correct_detection_psk}
    &\text{P}(\hat{\boldsymbol{s}}|\boldsymbol{x})\geq \displaystyle \frac{1}{2^K}\prod_{k=1}^K \text{erf}\pc{\frac{d_{1,k}}{\sigma_w}}+\text{erf}\pc{ \frac{d_{2,k}}{\sigma_w}}.
\end{align}
Based on \eqref{eq:Prob_correct_detection_psk} an optimization problem can be cast as
\begin{align}
\label{eq:opt_psk1}
    \displaystyle \arg\max_{\boldsymbol{x}\in \mathcal{X}^M}  &\displaystyle \prod_{k=1}^K \text{erf}\pc{\frac{d_{1,k}}{\sigma_w}}+\text{erf}\pc{ \frac{d_{2,k}}{\sigma_w}}.
\end{align}
Since $\log\pc{\cdot}$ is a monotonically increasing function, applying it to the objective from \eqref{eq:opt_psk1} yields an equivalent problem. With this, the union bound MSEP (UBMSEP) optimization problem for an $\alpha_s$-PSK modulation, reads as
\begin{align}
\label{eq:msep_prob_psk}
    \displaystyle \arg\min_{\boldsymbol{x}\in \mathcal{X}^M}  &\displaystyle -\sum_{k=1}^K \log \pc{\text{erf}\pc{\frac{d_{1,k}}{\sigma_w}}+\text{erf}\pc{ \frac{d_{2,k}}{\sigma_w}}} .
\end{align}
An alternative real valued formulation of the UBMSEP optimization problem can be cast as  
\begin{align}
\label{eq:rv_ubmsep_prob}
    &\displaystyle \min_{\boldsymbol{x}_\text{r}} \  -\sum_{k=1}^K \log \pc{ \text{erf}\pc{ \pc{\boldsymbol{h}_{\text{R},\theta,k}^{s^*} -\boldsymbol{h}_{\text{I},\theta,k}^{s^*}} \boldsymbol{x}_{\text{r}} }+
    \text{erf}\pc{ \pc{\boldsymbol{h}_{\text{R},\theta,k}^{s^*} +\boldsymbol{h}_{\text{I},\theta,k}^{s^*}} \boldsymbol{x}_{\text{r} } }} \\
   &\text{s.t.} \ \ \ \  {x}_{\text{r},2m-1}+j{x}_{\text{r},2m} \in \mathcal{X} \ \ \ \ \textrm{for } m=1,\ldots,M     \textrm{.} \notag
\end{align}
where $\boldsymbol{h}_{\text{R},\theta,k}^{s^*}$ and $\boldsymbol{h}_{\text{I},\theta,k}^{s^*}$ are the $k$-th rows of matrices $\boldsymbol{H}_{\text{R},\theta}^{s^*}$ and $\boldsymbol{H}_{\text{I},\theta}^{s^*}$, respectively, which are defined as $\boldsymbol{H}_{\text{R},\theta}^{s^*}=\frac{\sin\pc{\theta}}{\sigma_w}\boldsymbol{H}_{\text{R}}^{s^*}  $, and  $\boldsymbol{H}_{\text{I},\theta}^{s^*}=\frac{\cos\pc{\theta}}{\sigma_w} \boldsymbol{H}_{\text{I}}^{s^*}  $, with  
\begin{align}
&\quad\boldsymbol{H}_{\text{R}}^{s^*}    =      \begin{bmatrix} 
&\textrm{Re}\chav{{h}_{11}^{s^*}} &-\textrm{Im}\chav{{h}_{11}^{s^*}} 
                         &\cdots
&\textrm{Re}\chav{{h}_{1M}^{s^*}} \ &-\textrm{Im}\chav{{h}_{1M}^{s^*}}  \\
&\vdots &\vdots&\vdots&\vdots &\vdots&\\
&\textrm{Re}\chav{{h}_{K1}^{s^*}} &-\textrm{Im}\chav{{h}_{K1}^{s^*}} 
                         &\cdots
&\textrm{Re}\chav{{h}_{KM}^{s^*}} \ &-\textrm{Im}\chav{{h}_{KM}^{s^*}}  
\end{bmatrix}\\
&\quad\boldsymbol{H}_{\text{I}}^{s^*}    =      \begin{bmatrix} 
&\textrm{Im}\chav{{h}_{11}^{s^*}} &\textrm{Re}\chav{{h}_{11}^{s^*}} 
                         &\cdots
&\textrm{Im}\chav{{h}_{1M}^{s^*}} \ &\textrm{Re}\chav{{h}_{1M}^{s^*}}  \\
&\vdots &\vdots&\vdots&\vdots &\vdots&\\
&\textrm{Im}\chav{{h}_{K1}^{s^*}} &\textrm{Re}\chav{{h}_{K1}^{s^*}} 
                         &\cdots
&\textrm{Re}\chav{{h}_{KM}^{s^*}} \ &\textrm{Im}\chav{{h}_{KM}^{s^*}} 
\end{bmatrix},
\end{align}
where ${h}_{ij}^{s^*}$ is the element of the $i$-th row and $j$-th column of the matrix $\boldsymbol{H}^{s^*}=\text{diag}\chav{s^*} \boldsymbol{H}$.

%% file: figures/Union_bound.tex
\tikzset{every picture/.style={line width=0.75pt}} 

\begin{tikzpicture}[x=0.35pt,y=0.35pt,yscale=-1,xscale=1]

\draw    (236,421) -- (561,95) ;
\draw    (235.13,95.41) -- (559.13,418.41) ;
\draw  (152,257.41) -- (644,257.41)(398.13,23) -- (398.13,493) (637,252.41) -- (644,257.41) -- (637,262.41) (393.13,30) -- (398.13,23) -- (403.13,30)  ;
\draw  [color={rgb, 255:red, 65; green, 117; blue, 5 }  ,draw opacity=1 ][fill={rgb, 255:red, 65; green, 117; blue, 5 }  ,fill opacity=1 ] (438.75,142) -- (442.42,142) -- (442.42,144.75) -- (445.17,144.75) -- (445.17,148.58) -- (442.42,148.58) -- (442.42,151.33) -- (438.75,151.33) -- (438.75,148.58) -- (436,148.58) -- (436,144.75) -- (438.75,144.75) -- cycle ;
\draw  [color={rgb, 255:red, 255; green, 0; blue, 0 }  ,draw opacity=1 ][fill={rgb, 255:red, 252; green, 0; blue, 0 }  ,fill opacity=1 ] (507.58,210) -- (511.25,210) -- (511.25,212.75) -- (514,212.75) -- (514,216.58) -- (511.25,216.58) -- (511.25,219.33) -- (507.58,219.33) -- (507.58,216.58) -- (504.83,216.58) -- (504.83,212.75) -- (507.58,212.75) -- cycle ;
\draw  [color={rgb, 255:red, 65; green, 117; blue, 5 }  ,draw opacity=1 ][fill={rgb, 255:red, 65; green, 117; blue, 5 }  ,fill opacity=1 ] (506.75,296.67) -- (510.42,296.67) -- (510.42,299.42) -- (513.17,299.42) -- (513.17,303.25) -- (510.42,303.25) -- (510.42,306) -- (506.75,306) -- (506.75,303.25) -- (504,303.25) -- (504,299.42) -- (506.75,299.42) -- cycle ;
\draw  [color={rgb, 255:red, 0; green, 0; blue, 0 }  ,draw opacity=1 ][fill={rgb, 255:red, 0; green, 0; blue, 0 }  ,fill opacity=1 ] (541.58,182) -- (545.25,182) -- (545.25,184.75) -- (548,184.75) -- (548,188.58) -- (545.25,188.58) -- (545.25,191.33) -- (541.58,191.33) -- (541.58,188.58) -- (538.83,188.58) -- (538.83,184.75) -- (541.58,184.75) -- cycle ;
\draw    (506.47,152.36) -- (541.95,185.31) ;
\draw [shift={(543.42,186.67)}, rotate = 222.87] [color={rgb, 255:red, 0; green, 0; blue, 0 }  ][line width=0.75]    (10.93,-3.29) .. controls (6.95,-1.4) and (3.31,-0.3) .. (0,0) .. controls (3.31,0.3) and (6.95,1.4) .. (10.93,3.29)   ;
\draw [shift={(505,151)}, rotate = 42.87] [color={rgb, 255:red, 0; green, 0; blue, 0 }  ][line width=0.75]    (10.93,-3.29) .. controls (6.95,-1.4) and (3.31,-0.3) .. (0,0) .. controls (3.31,0.3) and (6.95,1.4) .. (10.93,3.29)   ;
\draw    (543.43,188.67) -- (543.98,256) ;
\draw [shift={(544,258)}, rotate = 269.53] [color={rgb, 255:red, 0; green, 0; blue, 0 }  ][line width=0.75]    (10.93,-3.29) .. controls (6.95,-1.4) and (3.31,-0.3) .. (0,0) .. controls (3.31,0.3) and (6.95,1.4) .. (10.93,3.29)   ;
\draw [shift={(543.42,186.67)}, rotate = 89.53] [color={rgb, 255:red, 0; green, 0; blue, 0 }  ][line width=0.75]    (10.93,-3.29) .. controls (6.95,-1.4) and (3.31,-0.3) .. (0,0) .. controls (3.31,0.3) and (6.95,1.4) .. (10.93,3.29)   ;
\draw  [draw opacity=0][fill={rgb, 255:red, 253; green, 0; blue, 0 }  ,fill opacity=0.1 ] (662.28,256.12) .. controls (662.28,256.12) and (662.28,256.12) .. (662.28,256.12) .. controls (662.28,256.12) and (662.28,256.12) .. (662.28,256.12) .. controls (662.28,389.33) and (544.35,497.32) .. (398.89,497.32) .. controls (330.55,497.32) and (268.29,473.49) .. (221.48,434.4) -- (398.89,256.12) -- cycle ;
\draw  [draw opacity=0][fill={rgb, 255:red, 80; green, 227; blue, 194 }  ,fill opacity=0.15 ] (135.11,258) .. controls (135.11,258) and (135.11,258) .. (135.11,258) .. controls (135.11,258) and (135.11,258) .. (135.11,258) .. controls (135.11,124.79) and (253.03,16.8) .. (398.5,16.8) .. controls (466.84,16.8) and (529.1,40.63) .. (575.91,79.72) -- (398.5,258) -- cycle ;
\draw  [draw opacity=0][fill={rgb, 255:red, 255; green, 0; blue, 0 }  ,fill opacity=0.1 ] (222.06,434.87) .. controls (168.07,391.42) and (134.11,328.37) .. (134.11,258.2) .. controls (134.11,257.93) and (134.11,257.65) .. (134.11,257.37) -- (398.78,258.2) -- cycle ;
\draw  [draw opacity=0][fill={rgb, 255:red, 80; green, 227; blue, 194 }  ,fill opacity=0.15 ] (223.05,435.5) .. controls (169.07,392.05) and (135.11,329) .. (135.11,258.83) .. controls (135.11,258.55) and (135.11,258.28) .. (135.11,258) -- (399.78,258.83) -- cycle ;

\draw (486,208.4)  node  [scale=0.7] [align=left] {$s_{i}$};
\draw (450,124.4) node  [scale=0.7] [align=left] {$s_{j_{1}}$};
\draw (535,302.82) node  [scale=0.7] [align=left] {$s_{j_{2}}$};
\draw (545,160.4) node  [scale=0.6] [align=left] {$d_{1}$};
\draw (572,220.4) node  [scale=0.6] [align=left] {$d_{2}$};

\end{tikzpicture}

%% file: Precoder/01_Precoding.tex
\input{Precoder/Suboptimal_Precoding}

\input{Precoder/Optimal_Precoding}

%% file: Precoder/Suboptimal_Precoding.tex
\section{MSEP Precoding using Projection Based Methods}
\label{sec:PBM_MSEP}

Projection based methods (PBMs) are prominent in literature due to its reduced computational complexity. Several works, e.g. \cite{MMSE_bb,MSM_precoder,Squid_precoder,CVX-CIO, park2021lowcomplexity,Sohrabi_msep_qam,nossek_vtc2021}, used PBMs with different design criteria. As such, in this section, different PBMs are considered to compute practical suboptimal solutions to \eqref{eq:rv_qmsep_prob} and \eqref{eq:reformulated_rv_ubmsep_prob}.

PBMs rely on the relaxation of the non-convex feasible set to a larger convex set, such that, the optimization problem can be easily solved with standard optimization problem tools. The solution of the relaxed problem is, then, projected into the original feasible set which yields a suboptimal solution to the problem. 

With this, we consider the relaxation of the discrete feasible set $\mathcal{X}^M$, to its convex hull $\mathcal{P}$. Note that the set $\mathcal{P}$ is a polyhedron and, thus, can described as the solution set of a finite number of linear equalities and inequalities \cite{Boyd_2004}. Similarly as done in \cite{MMSE_bb,MSM_precoder} and \cite{lopes2021discrete} the relaxed feasible set is described in real-valued notation using the inequality $\boldsymbol{R} \pr{ \boldsymbol{x}_{\text{r}}^T, 1}^T \leq \boldsymbol{0}$, where $\boldsymbol{R}=  \begin{bmatrix} \boldsymbol{A},   &-\boldsymbol{b} \end{bmatrix}$ and
\begin{align}
&\quad\boldsymbol{A}=\begin{bmatrix} (\boldsymbol{I}_M\otimes \boldsymbol{\beta}_1)^T, & (\boldsymbol{I}_M\otimes \boldsymbol{\beta}_2)^T,& \ldots, &(\boldsymbol{I}_M\otimes \boldsymbol{\beta}_{\alpha_x})^T \end{bmatrix}^T  \text{,} \\
&\quad \boldsymbol{\beta}_i=\begin{bmatrix} \cos{\phi_i},& - \sin{\phi_i}\end{bmatrix} ,
\quad  \phi_i=\frac{2\pi i }{\alpha_x}  \textrm{,  for  }  i=1,\ldots \textrm{,}\ \alpha_x  \textrm{,} \quad \boldsymbol{b}=\frac{\cos(\frac{\pi}{\alpha_x})}{\sqrt{{{M}}}} \boldsymbol{1}_{M\alpha_x} \text{.}
\end{align}
In the remainder of this study the projection step is denoted by the operator $\mathcal{M}(\cdot)$. The details of the different methods for implementing $\mathcal{M}(\cdot)$ will be detailed in subsection \ref{subsec:projection_methods}. In what follows the relaxed optimization problems are formulated for each MSEP design criterion.

\subsection{Relaxed Optimization Problems}
\label{subsec:relaxed_prob}
As mentioned before, a PBM's first step is to relax $\mathcal{X}^M$ to its convex hull $\mathcal{P}$. For the QMSEP case relaxing $\mathcal{X}^M$ to $\mathcal{P}$ leads to the following optimization problem
\begin{align}
\label{eq:qmsep_mapped_prob}
    &\min_{\boldsymbol{x}_\text{r}} \  -\sum_{k=1}^K \log \pc{\Phi\pc{\boldsymbol{h}_{\text{R},k} \boldsymbol{x}_\text{r}} }+ \log\pc{\Phi\pc{\boldsymbol{h}_{\text{I},k} \boldsymbol{x}_\text{r}}} \\
    &\text{s.t.} \ \ \ \  \boldsymbol{R} \pr{ \boldsymbol{x}_{\text{r}}^T, 1}^T \leq \boldsymbol{0}\notag.
\end{align}
Moreover, for the UBMSEP case the replacing $\mathcal{X}^M$ by $\mathcal{P}$ yields
\begin{align}
\label{eq:ubmsep_mapped_prob}
    &\displaystyle \min_{\boldsymbol{x}_\text{r}} \  -\sum_{k=1}^K \log \pc{ \text{erf}\pc{ \pc{\boldsymbol{h}_{\text{R},\theta,k}^{s^*} -\boldsymbol{h}_{\text{I},\theta,k}^{s^*}} \boldsymbol{x}_{\text{r}} }+
    \text{erf}\pc{ \pc{\boldsymbol{h}_{\text{R},\theta,k}^{s^*} +\boldsymbol{h}_{\text{I},\theta,k}^{s^*}} \boldsymbol{x}_{\text{r} } }} \\
   &\text{s.t.} \ \ \ \  \begin{bmatrix}  \boldsymbol{H}_{\text{R},\theta}^{s^*} - \boldsymbol{H}_{\text{I},\theta}^{s^*} \\ \boldsymbol{H}_{\text{R},\theta}^{s^*} + \boldsymbol{H}_{\text{I},\theta}^{s^*} 
\end{bmatrix} \boldsymbol{x}_\text{r} \geq \boldsymbol{0},   \ \ \ \ \boldsymbol{R} \pr{ \boldsymbol{x}_{\text{r}}^T, 1}^T \leq \boldsymbol{0}\notag.
\notag
\end{align}
Note that, replacing $\mathcal{X}^M$ by $\mathcal{P}$ yields convex problems since both \eqref{eq:qmsep_mapped_prob} and \eqref{eq:ubmsep_mapped_prob} minimize an convex objective under a convex feasible set.

\subsection{Projection Methods}
\label{subsec:projection_methods}
In the remainder of this work we denote the solution of the relaxed problems described in  \eqref{eq:qmsep_mapped_prob} and \eqref{eq:ubmsep_mapped_prob} as $\boldsymbol{x}_\text{r,lb}$ and its complex-valued description as $\boldsymbol{x}_\text{lb}$. Note that $\boldsymbol{x}_\text{lb}\in \mathcal{P}$ can also belong to the original feasible set  $\mathcal{X}^M$ as $\mathcal{P} \cap \mathcal{X}^M \neq \emptyset$. If this is the case $\boldsymbol{x}_\text{lb}$ is also the optimal solution from the original problem and the projection step can be skipped. However, if $\boldsymbol{x}_\text{lb} \notin \mathcal{X}^M$ the precoding vector is computed as $\boldsymbol{x}=\mathcal{M}(\boldsymbol{x}_\text{lb})$. In what follows two different approaches for implementing $\mathcal{M}(\cdot)$ are exposed.

\subsubsection{Projection Based on Uniform Quantization}
\label{subsec:mapped_msep}
$\\$
One of the most prominent projection methods present in literature is uniform quantization (UQ). When using this approach $\mathcal{M}(\cdot)=Q(\cdot)$, where $Q(\cdot)$ represents the quantization operation. The most common quantizing criterion is based on the elementwise Euclidean distance. As such, the $p$-th entry of the precoding vector $\boldsymbol{x}$, denoted as $x^p$, is computed as $x^p=\displaystyle\arg\hspace{-0.6em}\min_{i=1\hdots\alpha_x} \PM{x_\text{lb}^p-x_i}^2$, where $x_\text{lb}^p$ and $x^p$ denotes the $p$-th entry of $\boldsymbol{x}_\text{lb}$ and $\boldsymbol{x}$, respectively and $x_i$ the $i$-th element of $\mathcal{X}$.

\subsubsection{Projection via Greedy Search}
\label{subsec:GS_Msep}
$\\$
Although practical, UQ based projection can cause significant performance degradation.
As such, in this subsection, this performance degradation is mitigated using GS as a local optimization approach. 
GS algorithms compute for each entry of the quantized vector $\boldsymbol{x}_\text{ub}$ the value in $\mathcal{X}$ which yield the smallest objective $g(\boldsymbol{x})$. 
When using the QMSEP criterion $g(\boldsymbol{x})$ is given by 
\begin{align}
\label{eq:g_qmsep}
    g\pc{\boldsymbol{x}}= -\boldsymbol{1}_K \pc{\log \pc{\Phi\pc{\boldsymbol{S}_\text{r}\  \text{Re}\chav{\boldsymbol{Hx}}} }+\log \pc{\Phi\pc{\boldsymbol{S}_\text{i}\ \text{Im}\chav{\boldsymbol{Hx}}} }},
\end{align}
where $S_\text{r}=\frac{\sqrt{2}}{\sigma_w} \ \text{diag} \pc{\text{sign}(\text{Re}\chav{\boldsymbol{s}})}$ and $S_\text{i}=\frac{\sqrt{2}}{\sigma_w} \ \text{diag} \pc{\text{sign}(\text{Im}\chav{\boldsymbol{s}})}$. For the UBMSEP case $g(\boldsymbol{x})$ reads as 
\begin{align}
\label{eq:g_ubmsep}
    {g} (\boldsymbol{x})= -\boldsymbol{1}_K \log\pc{\text{erf}\pc{\text{Re}\chav{\boldsymbol{H}^\text{s} \boldsymbol{x}}-\text{Im}\chav{\boldsymbol{H}^\text{c}  \boldsymbol{x}}}+\text{erf}\pc{ \text{Re}\chav{\boldsymbol{H}^\text{s} \boldsymbol{x}}+\text{Im}\chav{\boldsymbol{H}^\text{c}  \boldsymbol{x}}}},
\end{align}
where $\boldsymbol{H}^{\text{s}}=\frac{\sin(\theta)}{\sigma_w} \text{diag}(\boldsymbol{s}^*)\boldsymbol{H}$ and $\boldsymbol{H}^{\text{c}}=\frac{\cos(\theta)}{\sigma_w} \text{diag}(\boldsymbol{s}^*)\boldsymbol{H} $. 
In this subsection two GS methods are considered for implementing $\mathcal{M}(\cdot)$. The first is a partial GS approach where only the entries of $\boldsymbol{x}_\text{lb}$ which do not belong to $\mathcal{X}$ are evaluated. The second is a full GS method where all entries of $\boldsymbol{x}_\text{lb}$ are considered. In the following, the steps of the methods are detailed. Algorithm \ref{alg:Partial_GS} exposes the partial GS approach while Algorithm \ref{alg:Full_GS} details the full GS method. Regarding notation $P$ denotes the length of the input vector.

\input{Precoder/Algorithms/Alg_GS_QMSEP}
\input{Precoder/Algorithms/Alg_GS_UBMSEP}


%% file: Precoder/Algorithms/Alg_GS_QMSEP.tex
\begin{algorithm}
\small
  \caption{Partial-GS Projection Algorithm}
	\label{alg:Partial_GS}
  \begin{algorithmic}    
  \State{\textbf{Inputs}: $\boldsymbol{\boldsymbol{x}_\text{ub}}$, $\boldsymbol{\boldsymbol{x}_\text{lb}}$ and $g(\boldsymbol{x})$ \hspace{1em} \textbf{Output}: $\boldsymbol{x}_\text{out}$ }
  \State{Construct the set $\mathcal{T}= \chav{p: x_\text{lb}^p \notin \mathcal{X} }$  }
    \For{$p \in \mathcal{T}$}
	  \For{$i=1:\alpha_x$}
		\State{Fix ${x}^p_\text{ub}$ as $x_i$ and compute the $g_{p}^i=g(\boldsymbol{x}_\text{ub})$}
	\EndFor
    \State{Update the $p$-th entry of $\boldsymbol{x}_\text{ub}$ as $x^p_\text{ub}=\displaystyle \arg\hspace{-0.9em}\min_{i=1,\hdots,\alpha_x} g_p^i$ }
	\EndFor
	\State{The output vector is given by $\boldsymbol{x}_\text{out}=\boldsymbol{x}_\text{ub}$}
\end{algorithmic}
\end{algorithm}

%% file: Precoder/Algorithms/Alg_GS_UBMSEP.tex
\begin{algorithm}
\small
  \caption{Full-GS Projection Algorithm}
	\label{alg:Full_GS}
  \begin{algorithmic}    
  \State{\textbf{Inputs}: $\boldsymbol{\boldsymbol{x}_\text{ub}}$, $g(\boldsymbol{x})$ and $P$  \hspace{1em} \textbf{Output}: $\boldsymbol{x}_\text{out}$ }
    \For{$p=1 : P$}
	  \For{$i=1:\alpha_x$}
		\State{Fix ${x}^p_\text{ub}$ as $x_i$ and compute the $g_{p}^i=g(\boldsymbol{x}_\text{ub})$}
	\EndFor
    \State{Update the $p$-th entry of $\boldsymbol{x}_\text{ub}$ as $x^p_\text{ub}=\displaystyle \arg\hspace{-0.9em}\min_{i=1,\hdots,\alpha_x} g_p^i$ }
    \EndFor
	\State{The output vector is given by $\boldsymbol{x}_\text{out}=\boldsymbol{x}_\text{ub}$}
\end{algorithmic}
\end{algorithm}

%% file: Precoder/Optimal_Precoding.tex
\section{Optimal MSEP Precoding via Branch-and-Bound}
\label{sec:MSEP_BB}

In this section, we devise algorithms that solve optimally the DPPs described in \eqref{eq:rv_qmsep_prob} and \eqref{eq:reformulated_rv_ubmsep_prob}. According to \cite{Belenky1998}, the most established class of approaches for solving a DPP with reasonable computational complexity is the B\&B.
First created in 1960 by A. H. Land and A. G. Doig \cite {Bnb_invention} the essence of B\&B methods consists in eliminating, as many as possible, candidate solutions, such that, the optimal can be computed using exhaustive search. The basics of precoding using B\&B method can be found well detailed in the works from \cite{General_MMDDT_BB,Landau2017,MMSE_bb,lopes2021discrete}, as such, this study will focus on the specifics of the proposed B\&B algorithms.

\subsection{Branch-and-Bound as a Tree Search Method}
\label{subsec:bb_tree}
A B\&B algorithm is a tree search based method where the tree represents the feasible set. For the case of the problems defined in \eqref{eq:rv_qmsep_prob} and \eqref{eq:reformulated_rv_ubmsep_prob} the feasible set of the optimization variable is $\mathcal{X}^M$. As such, in this subsection we show how $\mathcal{X}^M$ is represented in the form of a tree. 

\begin{figure} 
\centering
\input{figures/Tree}
\caption{Tree representation of the set $\mathcal{X}^M$ for a system with $M=2$ BS antennas and QPSK precoding modulation ($\alpha_x=4$)}
\label{fig:tree}       
\end{figure}
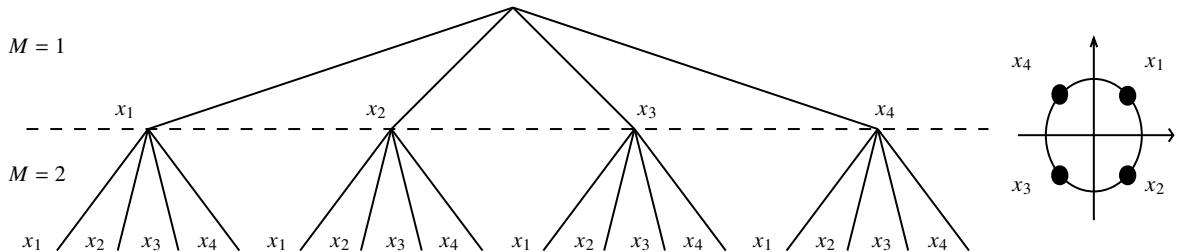

For the construction of the tree it is considered that the $p$-th BS antenna represents the $p$-th layer and each possibility for a subvector $\boldsymbol{f}\in \mathcal{X}^p$ represents one branch. 
With this, the tree has $M$ layers with $\alpha_x^M$ branches in the last layer. An example of a tree representation of the feasible set is shown in Fig.~\ref{fig:tree} for the case of two transmit antennas at the BS and QPSK signaling.


\subsection{Subproblem formulation}

In a B\&B algorithm a DPP is solved by considering partially fixed subvectors and computing upper and lower bounds to evaluate if the fixed subvector is part of the optimal solution. In this subsection, we describe, for each different MSEP formulation, the optimization problems that are solved in each layer of the B\&B algorithm to compute the lower bounding step. 

As mentioned in subsection \ref{subsec:bb_tree}, the branches of the tree represent a subvector $\boldsymbol{f} \in \mathcal{X}^p$ for the $p$-th layer. With this, the subproblems are derived by fixing, for each branch, the corresponding subvector $\boldsymbol{f}$ and optimizing the remaining subvector $\boldsymbol{v}\in\mathcal{X}^{M-p}$, as such the transmit vector is given by $\boldsymbol{x}=\pr{\boldsymbol{f}^T , \ \boldsymbol{v}^T}^T$.
However, such that the optimization problems are real-valued, in this study it is considered the division of $\boldsymbol{x}_\text{r}$ instead of $\boldsymbol{x}$, which reads as
\begin{align}
    \boldsymbol{x}_{\text{r}}    =      \begin{bmatrix} 
\boldsymbol{f}_{\text{r}}^T , \ \boldsymbol{v}_{\text{r}}^T
\end{bmatrix}^T,
\end{align}
where for the $p$-th layer, the length of the fixed subvector $\boldsymbol{f}_\text{r}$ is $2 p$ and, consequently, the length of the subvector $\boldsymbol{v}_\text{r}$ is $2(M-p)$. The subproblems are then derived based on the formulation of the relaxed PBM problems from subsection \ref{subsec:relaxed_prob}.

\subsubsection{QMSEP Subproblem formulation}
$\\$
The QMSEP subproblems are written considering the minimization of the objective function shown in \eqref{eq:g_qmsep} for a given $\boldsymbol{f}_\text{r}$. To this end, the matrices $\boldsymbol{H}_{\text{R}}$ and $\boldsymbol{H}_{\text{I}}$ are divided as
\begin{align}
\quad\boldsymbol{H}_{\text{R}}   =      \begin{bmatrix} 
\boldsymbol{F}_{\text{R}}\ , \ \boldsymbol{V}_{\text{R}}
\end{bmatrix}
\quad\boldsymbol{H}_{\text{I}}   =      \begin{bmatrix} 
\boldsymbol{F}_{\text{I}}\ , \ \boldsymbol{V}_{\text{I}}
\end{bmatrix}
\end{align}
where $\boldsymbol{F}_{\text{R}}$ and $\boldsymbol{F}_{\text{I}}$ consist of the first $2p$ columns of $\boldsymbol{H}_{\text{R}}$ and $\boldsymbol{H}_{\text{I}}$, respectively and $\boldsymbol{V}_{\text{R}}$ and $\boldsymbol{V}_{\text{I}}$ consist of the subsequent $2(M-p)$ columns of $\boldsymbol{H}_{\text{R}}$ and $\boldsymbol{H}_{\text{I}}$, respectively.
With this, the subproblem conditioned on $\boldsymbol{f}_\text{r}$ reads as
\begin{align}
\label{eq:qpsk_sub_prob}
    &\min_{\boldsymbol{v}_\text{r}} \  -\boldsymbol{1}_K \pc{\log \pc{\Phi\pc{\boldsymbol{F}_\text{R} \boldsymbol{f}_\text{r}+\boldsymbol{V}_\text{R} \boldsymbol{v}_\text{r}} } + \log \pc{\Phi\pc{\boldsymbol{F}_\text{I} \boldsymbol{f}_\text{r}+\boldsymbol{V}_\text{I} \boldsymbol{v}_\text{r}} }} \\
    &\text{s.t.} \ \ \ \  \boldsymbol{R}^\prime
\pr{ \boldsymbol{v}_{\text{r}}^T, 1}^T\leq \boldsymbol{0} \notag,
\end{align}
where $\boldsymbol{R}^{\prime}=\pr{ \boldsymbol{A}^{\prime},\  -\boldsymbol{b}^{\prime}  }$ is obtained by selecting the last $2\pc{M-p}$ columns of $\boldsymbol{R}$. 

\subsubsection{UBMSEP Subproblem formulation}
$\\$
Similarly as in the QMSEP case, the matrices $\boldsymbol{H}_{\text{R}}^{s^*}$ and $\boldsymbol{H}_{\text{I}}^{s^*}$ are divided as
\begin{align}
\quad\boldsymbol{H}_{\text{R}}^{s^*}    =      \begin{bmatrix} 
\boldsymbol{F}_{\text{R},\theta}^{s^*} \ ,\ \boldsymbol{V}_{\text{R},\theta}^{s^*}
\end{bmatrix}
\quad\boldsymbol{H}_{\text{I}}^{s^*}    =      \begin{bmatrix} 
\boldsymbol{F}_{\text{I},\theta}^{s^*} \ ,\ \boldsymbol{V}_{\text{I},\theta}^{s^*}
\end{bmatrix}.
\end{align}
As such, the subproblem associated with the fixed vector $\boldsymbol{f}_\text{r}$ is given by
\begin{align}
\label{eq:PSK_sub_problem_relaxed}
&\displaystyle \min_{\boldsymbol{v}_\text{r}}  -\boldsymbol{1}_K \pc{ \log \pc{    \text{erf}\pc{ \boldsymbol{F}_{\text{R},\theta}^{s^*}\boldsymbol{f}_{\text{r}} + \boldsymbol{V}_{\text{R},\theta}^{s^*}\boldsymbol{v}_{\text{r}} -\boldsymbol{F}_{\text{I},\theta}^{s^*}\boldsymbol{f}_{\text{r}}+\boldsymbol{V}_{\text{I},\theta}^{s^*}\boldsymbol{v}_{\text{r}} }+
    \text{erf}\pc{ \boldsymbol{F}_{\text{R},\theta}^{s^*}\boldsymbol{f}_{\text{r}} + \boldsymbol{V}_{\text{R},\theta}^{s^*}\boldsymbol{v}_{\text{r}} +\boldsymbol{F}_{\text{I},\theta}^{s^*}\boldsymbol{f}_{\text{r}}+\boldsymbol{V}_{\text{I},\theta}^{s^*}\boldsymbol{v}_{\text{r}}} } }\notag \\
    &\text{s.t.} \ \ \ \ \begin{bmatrix}  \boldsymbol{V}_{\text{R},\theta}^{s^*}-\boldsymbol{V}_{\text{I},\theta}^{s^*} \\ \boldsymbol{V}_{\text{R},\theta}^{s^*}+\boldsymbol{V}_{\text{I},\theta}^{s^*} 
\end{bmatrix} \boldsymbol{v}_\text{r} \geq -\begin{bmatrix}  \boldsymbol{F}_{\text{R},\theta}^{s^*} - \boldsymbol{F}_{\text{I},\theta}^{s^*} \\ \boldsymbol{F}_{\text{R},\theta}^{s^*} + \boldsymbol{F}_{\text{I},\theta}^{s^*} 
\end{bmatrix}\boldsymbol{f}_{\text{r}} ,   \ \ \ \  \boldsymbol{R}^\prime
\pr{ \boldsymbol{v}_{\text{r}}^T, 1}^T\leq \boldsymbol{0}.
\end{align}

\subsection{Proposed MSEP Branch-and-Bound Algorithms}

In this subsection a B\&B algorithm is assembled with the tools previously presented. Utilizing the B\&B algorithm requires choosing a MSEP criterion and a projection method. As such, it is considered $\mathcal{M}(\cdot)$, the criterion utilized and consequently $g(\boldsymbol{x})$ as inputs of the algorithm.

\subsubsection{Initialization Step}
$\\$
In the proposed MSEP B\&B algorithms an initialization step is considered. As such, the relaxed problem, described in subsection \ref{subsec:relaxed_prob}, corresponding to the chosen criterion is solved which yields $\boldsymbol{x}_\text{r,lb}$. Subsequently, the projection step is performed which yields $\check{\boldsymbol{x}}=\mathcal{M}(C(\boldsymbol{x}_\text{r,lb}))$. 
Note that, as mentioned in \cite{MMSE_bb} and \cite{lopes2021discrete}, if $\boldsymbol{x}_{\text{lb}}=\check{\boldsymbol{x}}$ the algorithm returns $\check{\boldsymbol{x}}$ as it is the optimal solution. Otherwise, $\check{g}=g(\check{\boldsymbol{x}})$ is computed as the initial smallest known upper bound and both $\check{g}$ and $\check{\boldsymbol{x}}$ are stored.

\subsubsection{Tree search process}
$\\$
For the tree search process breadth first search is considered. The process starts by setting the layer value $p=1$ and, accordingly, solving the subproblems which yields the solution $\boldsymbol{v}_{\text{r,lb}|\boldsymbol{f}}$. The vector $\boldsymbol{x}_{\text{lb}|\boldsymbol{f}}=\pr{\boldsymbol{f}^T, C\pc{\boldsymbol{v}_{\text{r,lb}|\boldsymbol{f}}}^T }^T$ is, then, constructed and the value of $g(\boldsymbol{x}_{\text{lb}|\boldsymbol{f}})$ is computed and stored. 
The solution subvector $\boldsymbol{v}_{\text{lb}|\boldsymbol{f}}$ is projected to $\mathcal{X}^{M-p}$ which yields $\boldsymbol{v}_{\text{ub}|\boldsymbol{f}}=\mathcal{M}(\boldsymbol{v}_{\text{lb}|\boldsymbol{f}})$. With this, one can construct $\boldsymbol{x}_{\text{ub}|\boldsymbol{f}}=\pr{\boldsymbol{f}^T, \boldsymbol{v}_{\text{ub}|\boldsymbol{f}}^T}^T$. Note that, $\boldsymbol{x}_{\text{ub}|\boldsymbol{f}} \in \mathcal{X}^{M}$ is an upper bound solution, and thus, $g(\boldsymbol{x}_{\text{ub}|\boldsymbol{f}})$ is an upper bound on $g(\boldsymbol{x}_{\text{opt}})$, with $\boldsymbol{x}_{\text{opt}}$ being the optimal solution.

To evaluate if $g(\boldsymbol{x}_{\text{ub}|\boldsymbol{f}})$ is the smallest known upper bound the condition $g(\boldsymbol{x}_{\text{ub}|\boldsymbol{f}})<\check{g}$ is checked. If true, the smallest known upper bound and its corresponding value of $\boldsymbol{x}$ are updated as $\check{g}=g(\boldsymbol{x}_{\text{ub}|\boldsymbol{f}})$ and $\check{\boldsymbol{x}}=\boldsymbol{x}_{\text{ub}|\boldsymbol{f}}$. 

After all possible valid branches in one layer are evaluated, i.e. all valid values of $\boldsymbol{f}$ were fixed and its conditioned upper and lower bounds computed, they are considered in the pruning process. The pruning step is explained in details in the following subsection. After pruning, the set of valid $\boldsymbol{f}$ subvectors is updated and the algorithm repeats this process in the next layer.
In the last layer, it is expected that only a few valid candidate solutions remain. As such, they are all evaluated against $\check{\boldsymbol{x}}$ and the optimal value is determined by the vector that yields the minimum value of the objective function. Note that the optimal solution might not be in the last layer as it could be found in previous layers. 

\subsubsection{Pruning Step}
$\\$
To determine if a fixed subvector $\boldsymbol{f}$ is valid, i.e. if it can be a subvector of the optimal solution $\boldsymbol{x}_\text{opt}$, the stored values $g(\boldsymbol{x}_{\text{lb}|\boldsymbol{f}})$ are compared with $\check{g}$. In this work, a subvector is considered as valid if $g(\boldsymbol{x}_{\text{lb}|\boldsymbol{f}}) < (1-\gamma) \check{g}$, where $0 < \gamma \leq 1$, which yields a solution in the $\epsilon$-suboptimal set.

When evaluating the $p$-th layer for a branch, which is the same as evaluating the $p$-th layer for a subvector $\boldsymbol{f}$, it holds that $g(\boldsymbol{x}_{\text{lb}|\boldsymbol{f}}) \leq g(\boldsymbol{x}_{\text{ub}|\boldsymbol{f}})$. 
Consider now the scenario where $\boldsymbol{x}_{\text{ub}|\boldsymbol{f}}=\boldsymbol{x}_\text{opt}$ and  $g(\boldsymbol{x}_\text{opt}) = g(\boldsymbol{x}_{\text{lb}|\boldsymbol{f}})$. 
In this case, using the mentioned pruning condition, the optimal branch is considered as not valid if $g(\boldsymbol{x}_\text{opt}) \geq (1-\gamma) \check{g}$. This results in the algorithm returning $\check{\boldsymbol{x}}$ when $g(\boldsymbol{x}_\text{opt})\leq\check{g}\leq \frac{g(\boldsymbol{x}_\text{opt})}{1-\gamma}$. This means that the algorithm allows for any solution in the $\epsilon$-suboptimal set given by $\mathcal{X}_{\text{opt},\epsilon}=\chav{\boldsymbol{x}: \  \text{g}(\boldsymbol{x}) \leq g(\boldsymbol{x}_\text{opt})+ \epsilon}$, cf.\ \cite{Boyd_2004}, where $\epsilon=g(\boldsymbol{x}_\text{opt})\frac{\gamma}{1-\gamma}$.

Allowing for a transmit vector in the $\epsilon$-suboptimal theoretically could yield a suboptimal solution. However, due to the discrete nature of the feasible set, one can choose a sufficiently small value for $\gamma$, for example $\gamma=\delta \check{g}$ which yields $\epsilon=g(\boldsymbol{x}_\text{opt})\frac{\delta \check{g}}{1-\delta \check{g}}$ with $0 \leq \delta \ll 1$, such that $\mathcal{X}_{\text{opt},\epsilon}$ contains only the global optimal solution $\boldsymbol{x}_\text{opt}$.

By setting a sufficiently small value for $\delta$, the output solution of the proposed B\&B algorithm corresponds, with probability one, to the optimal solution $\boldsymbol{x}_\text{opt}$ of the corresponding MSEP problem, as is confirmed by the numerical results shown in Section~\ref{sec:numerical_results}. The steps of the MSEP B\&B algorithm are detailed in Algorithm~\ref{alg:MSEP}.

\input{Precoder/Algorithms/Alg_MSEP}

%% file: figures/Tree.tex
\tikzset{every picture/.style={line width=0.75pt}} 

\begin{tikzpicture}[x=0.25pt,y=0.5pt,yscale=-1,xscale=1]

\draw    (736,46) -- (1288,138) ;
\draw    (736,46) -- (184,138) ;
\draw    (736,46) -- (920,138) ;
\draw    (736,46) -- (552,138) ;
\draw    (184,138) -- (322,230) ;
\draw    (184,138) -- (230,230) ;
\draw    (184,138) -- (138,230) ;
\draw    (184,138) -- (46,230) ;
\draw    (552,138) -- (690,230) ;
\draw    (552,138) -- (598,230) ;
\draw    (552,138) -- (506,230) ;
\draw    (552,138) -- (414,230) ;
\draw    (920,138) -- (1058,230) ;
\draw    (920,138) -- (966,230) ;
\draw    (920,138) -- (874,230) ;
\draw    (920,138) -- (782,230) ;
\draw    (1288,138) -- (1426,230) ;
\draw    (1288,138) -- (1334,230) ;
\draw    (1288,138) -- (1242,230) ;
\draw    (1288,138) -- (1150,230) ;
\draw  [dash pattern={on 4.5pt off 4.5pt}]  (0,138) -- (309.5,138) -- (1455.5,138) ;
\draw  (1500,142.65) -- (1735.22,142.65)(1614.89,69) -- (1614.89,207) (1728.22,137.65) -- (1735.22,142.65) -- (1728.22,147.65) (1609.89,76) -- (1614.89,69) -- (1619.89,76)  ;
\draw   (1542.29,142.65) .. controls (1542.29,119.13) and (1574.79,100.06) .. (1614.89,100.06) .. controls (1654.99,100.06) and (1687.49,119.13) .. (1687.49,142.65) .. controls (1687.49,166.18) and (1654.99,185.25) .. (1614.89,185.25) .. controls (1574.79,185.25) and (1542.29,166.18) .. (1542.29,142.65) -- cycle ;
\draw  [color={rgb, 255:red, 0; green, 0; blue, 0 }  ,draw opacity=1 ][fill={rgb, 255:red, 0; green, 0; blue, 0 }  ,fill opacity=1 ] (1657.83,107.96) .. controls (1662.37,105.3) and (1669.73,105.3) .. (1674.27,107.96) .. controls (1678.81,110.63) and (1678.81,114.94) .. (1674.27,117.6) .. controls (1669.73,120.26) and (1662.37,120.26) .. (1657.83,117.6) .. controls (1653.3,114.94) and (1653.3,110.63) .. (1657.83,107.96) -- cycle ;
\draw  [color={rgb, 255:red, 0; green, 0; blue, 0 }  ,draw opacity=1 ][fill={rgb, 255:red, 0; green, 0; blue, 0 }  ,fill opacity=1 ] (1555.16,168.2) .. controls (1559.7,165.54) and (1567.06,165.54) .. (1571.6,168.2) .. controls (1576.14,170.86) and (1576.14,175.18) .. (1571.6,177.84) .. controls (1567.06,180.5) and (1559.7,180.5) .. (1555.16,177.84) .. controls (1550.63,175.18) and (1550.63,170.86) .. (1555.16,168.2) -- cycle ;
\draw  [color={rgb, 255:red, 0; green, 0; blue, 0 }  ,draw opacity=1 ][fill={rgb, 255:red, 0; green, 0; blue, 0 }  ,fill opacity=1 ] (1555.17,107) .. controls (1559.39,104.15) and (1566.48,104) .. (1571.02,106.66) .. controls (1575.56,109.32) and (1575.81,113.79) .. (1571.6,116.64) .. controls (1567.38,119.48) and (1560.28,119.64) .. (1555.74,116.98) .. controls (1551.2,114.32) and (1550.95,109.85) .. (1555.17,107) -- cycle ;
\draw  [color={rgb, 255:red, 0; green, 0; blue, 0 }  ,draw opacity=1 ][fill={rgb, 255:red, 0; green, 0; blue, 0 }  ,fill opacity=1 ] (1657.83,168.2) .. controls (1662.37,165.54) and (1669.73,165.54) .. (1674.27,168.2) .. controls (1678.81,170.86) and (1678.81,175.18) .. (1674.27,177.84) .. controls (1669.73,180.5) and (1662.37,180.5) .. (1657.83,177.84) .. controls (1653.3,175.18) and (1653.3,170.86) .. (1657.83,168.2) -- cycle ;

\draw (150,125) node  [scale=0.7] [align=left]    {$x_{1}$};
\draw (530,125) node  [scale=0.7] [align=left]    {$x_{2}$};
\draw (940,125) node  [scale=0.7] [align=left]    {$x_{3}$};
\draw (1300,125) node  [scale=0.7] [align=left]    {$x_{4}$};
\draw (10,225) node  [scale=0.7] [align=left]    {$x_{1}$};
\draw (105,225) node  [scale=0.7] [align=left]    {$x_{2}$};
\draw (190,225) node  [scale=0.7] [align=left]    {$x_{3}$};
\draw (275,225) node  [scale=0.7] [align=left]    {$x_{4}$};
\draw (380,225) node  [scale=0.7] [align=left]    {$x_{1}$};
\draw (475,225) node  [scale=0.7] [align=left]    {$x_{2}$};
\draw (560,225) node  [scale=0.7] [align=left]    {$x_{3}$};
\draw (640,225) node  [scale=0.7] [align=left]    {$x_{4}$};
\draw (750,225) node  [scale=0.7] [align=left]    {$x_{1}$};
\draw (845,225) node  [scale=0.7] [align=left]    {$x_{2}$};
\draw (930,225) node  [scale=0.7] [align=left]    {$x_{3}$};
\draw (1010,225) node  [scale=0.7] [align=left]    {$x_{4}$};
\draw (1115,225) node  [scale=0.7] [align=left]    {$x_{1}$};
\draw (1210,225) node  [scale=0.7] [align=left]    {$x_{2}$};
\draw (1295,225) node  [scale=0.7] [align=left]    {$x_{3}$};
\draw (1380,225) node  [scale=0.7] [align=left]    {$x_{4}$};
\draw (16,74.4) node  [scale=0.7] [align=left]    {$M=1$};
\draw (16,171.4) node  [scale=0.7] [align=left]    {$M=2$};
\draw (1708.01,88.4) node [scale=0.7] [align=left]    {$x_{1}$};
\draw (1708.01,182.4) node [scale=0.7] [align=left]    {$x_{2}$};
\draw (1506.73,182.4) node [scale=0.7] [align=left]    {$x_{3}$};
\draw (1506.73,88.4) node [scale=0.7] [align=left]    {$x_{4}$};

\end{tikzpicture}

%% file: Precoder/Algorithms/Alg_MSEP.tex
\begin{algorithm}
\small
  \caption{Proposed MSEP B\&B Precoding Algorithm}
	\label{alg:MSEP}
  \begin{algorithmic}    
  \State{\textbf{Inputs}: $\mathcal{M}(\cdot)$, $g(\boldsymbol{x})$ and MSEP criterion \hspace{1em} \textbf{Output}: $\boldsymbol{x}_\text{out}$ }
    \State{Solve the relaxed optimization problem corresponding to the MSEP criterion and get $\boldsymbol{x}_\text{r,lb}$}
    \State{Project $\boldsymbol{x}_\text{lb}=R(\boldsymbol{x}_\text{r,lb})$ to get the upper bound solution $\boldsymbol{x}_\text{ub}=\mathcal{M}(\boldsymbol{x}_\text{lb})$ }
    \State{\textbf{If} $\boldsymbol{x}_\text{ub}==\boldsymbol{x}_\text{lb}$}
    \vspace{-0.5em}
    \State{\hspace{1em} \textbf{return} $\boldsymbol{x}_\text{out}=\boldsymbol{x}_\text{ub}$ }
    \vspace{-0.5em}
    \State{\textbf{end if} }
    \State{Define $\check{\boldsymbol{x}}=\boldsymbol{x}_\text{ub}$ and compute $\check{g}=g(\check{\boldsymbol{x}})$ }
    \State{Define the first level ($p=1$) of the tree by $\mathcal{G}_{p}:=\mathcal{X}_{\textrm{}}$}
	\For{$p=1:M-1$}
	\State{Partition  $\mathcal{G}_{p}$ in $\boldsymbol{f}_{1},\ldots,\boldsymbol{f}_{\PM{\mathcal{G}_{p}}}$ }  
	  \For{$i=1:\left| \mathcal{G}_{p} \right|$}
		\State{Conditioned on $\boldsymbol{f}_{\text{r},i}=R(\boldsymbol{f}_{i})$ solve the subproblem corresponding to the MSEP criterion to get $\boldsymbol{v}_{\text{r,lb}|f_{i}}$}
		\State{Construct $\boldsymbol{x}_{\text{lb,}i}=\pr{\boldsymbol{f}_{i}^T\ , \ C\pc{\boldsymbol{v}_{\text{r,lb}|f_{i}} }^T}$ and determine the lower bound $g_{\text{lb},i}=g(\boldsymbol{x}_{\text{lb,}i})$ }
		\State{Compute the upper bound $\boldsymbol{x}_{\text{ub,}i}=\mathcal{M}(\boldsymbol{x}_{\text{lb,}i})$}
		\State{With $\boldsymbol{x}_{\textrm{ub,i}}$ compute the upper bound $g_{\text{ub},i}=g(\boldsymbol{x}_{\text{ub,}i})$}
		\State{Update the best upper bound as $\check{g} =\min\left( \check{g}, g_{\text{ub},i}  \right)$ and update $\check{\boldsymbol{x}}$ accordingly} 
	\EndFor
	\State{Construct a reduced set by comparing conditioned  lower bounds with the global upper bound $\check{g}$}
	\State{$\mathcal{G}_{p}^{\prime}:=\left\{  \boldsymbol{x}_{\text{lb},i}\ \vert\  g_{\textrm{lb},i} < (1-\gamma) \ \check{g}       , i=1,\ldots,  \left|\mathcal{G}_{p}\right|  \right\} $}
	\State{Define the set for the next level in the tree: $\mathcal{G}_{p+1}:=\mathcal{G}_{p}^{\prime} \times \mathcal{X}_{\textrm{}}$}
	\EndFor
	\State{The global solution is $\boldsymbol{x}_{\textrm{out}} = \mathrm{arg} \hspace{-6mm}\displaystyle \min_{\boldsymbol{x} \in \chav{\mathcal{G}_M \cup \chav{\check{\boldsymbol{x}}}}  } g (\boldsymbol{x})$ }
\end{algorithmic}
\end{algorithm}

%% file: Numerical_results/01_Numerical_Results.tex
\section{Numerical Results}
\label{sec:numerical_results}

For the numerical evaluation, the SER is considered. We assume that the channel gains are modeled by independent Rayleigh fading \cite{Marzetta_2013}, meaning $\beta_m=1 \ \text{for} \ m=1,...,M$ and $g_{k,m}\sim \mathcal{CN}  ({0},\sigma_g^2)\ \text{for} \ k=1,...,K \ \text{and}\ m=1, ...,M$ as done implicitly in \cite{ZF-Precoding} and \cite{MSM_precoder} and explicitly in \cite{CVX-CIO}. Moreover, the SNR is defined by $\mathrm{SNR}=\frac{ \left\|\boldsymbol{x}\right\|^2_2 }{N_0}$, where the spectral noise power density $N_0$ is equivalent to the noise sample variance $\sigma_w^2$. For the MSEP B\&B methods $\delta$ is set to $\delta=5 \cdot 10^{-7}$.

This section is divided in three parts. In the first, the proposed MSEP formulations are evaluated against other state-of-the-art criteria. In the second, the proposed methods are compared with other state-of-the-art algorithms in terms of SER. Finally, in the third, a complexity analysis is performed and the proposed approaches are compared with the state-of-the-art techniques both in terms of runtime. 

The state-of-the-art methods considered for comparison in this study are 
\begin{itemize}
    \item[1.] The MSM-Precoder \cite{MSM_precoder} considering phase quantization;
    \item[2.] The ZF precoder with constant envelope \cite{ZF-Precoding}, where the entries of the precoding vector are subsequently phase quantized;
    \item[3.] The phase quantized CIO precoder implemented via CVX \cite{CVX-CIO};
    \item[4.] {The single-carrier version of the SQUID-OFDM precoder \cite{jacobsson2018nonlinear} phase quantized};
    \item[6.] {The C3PO precoder \cite{Struder_c3po} phase quantized considering a real valued scaling factor};
    \item[7.] The MMDDT branch-and-bound precoder \cite{General_MMDDT_BB};
    \item[8.] The MMSE branch-and-bound preocder \cite{MMSE_bb};
\end{itemize}

\subsection{Criteria comparison}
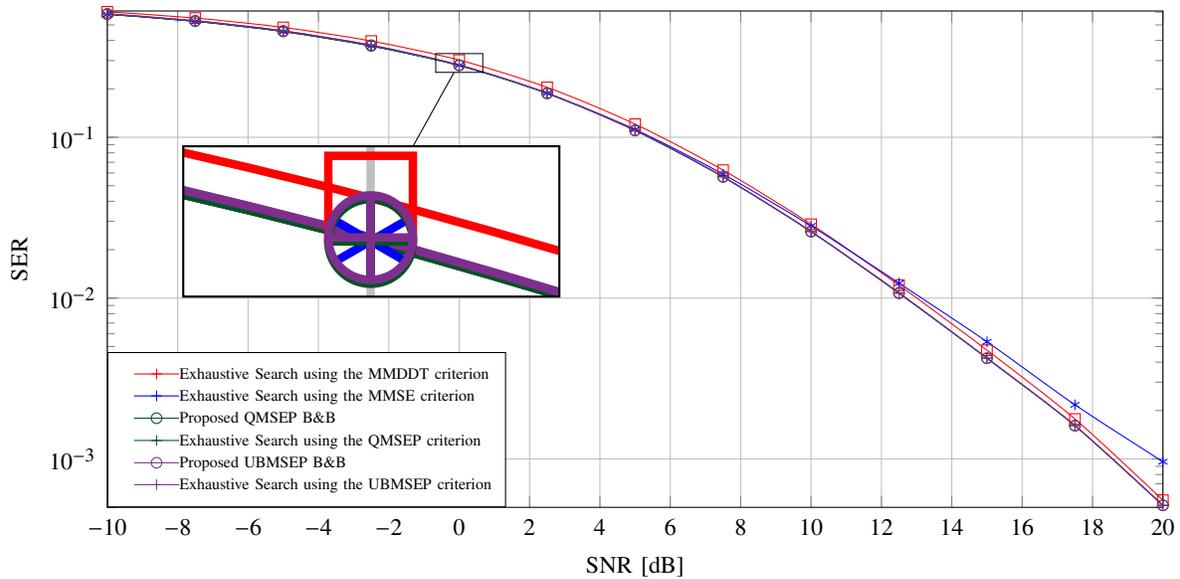
\begin{figure}[t]
\begin{center}
\input{Numerical_results/Figures/Uncoded_BB_Precoders}
\caption{SER versus $\mathrm{SNR}$ for $K=2$, $M=5$, $\alpha_s=4$ and $\alpha_x=4$} 
\label{fig:crit_comp_q}       
\end{center}
\end{figure}

\begin{figure}[t]
\begin{center}
\input{Numerical_results/Figures/Uncoded_BB}
\caption{SER versus $\mathrm{SNR}$ for $K=2$, $M=5$, $\alpha_s=8$ and $\alpha_x=8$} 
\label{fig:crit_comp_8}       
\end{center}
\end{figure}
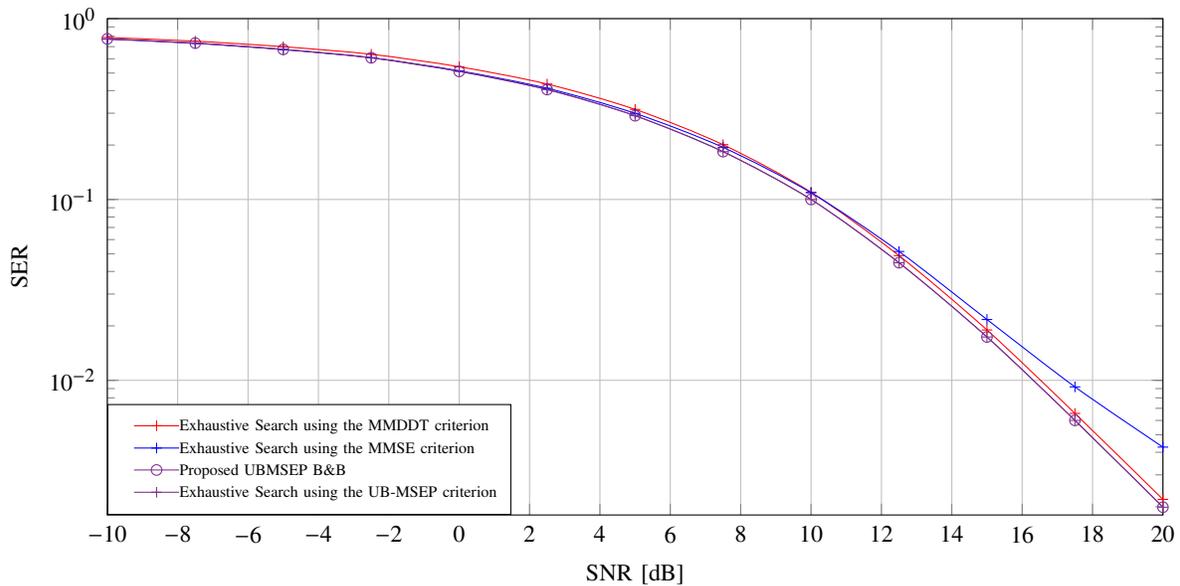

For the criteria comparison we consider two different scenarios. In the first, both MSEP criteria are compared with the MMSE and MMDDT criteria using QPSK data symbols. In the second, only the UBMSEP criterion is compared with the MMSE and MMDDT criterion using 8-PSK data symbols.

As can be seen in Fig.~\ref{fig:crit_comp_q}, the QMSEP criterion outperforms all other state-of-the-art criteria for most of the SNR range, having the same performance as MMSE for low-SNR regime. The UBMSEP method presents does not present significant decrease in performance when compared with the QMSEP approach. 
As shown in Fig.~\ref{fig:crit_comp_8}, the UBMSEP criterion also outperforms in terms of SER the state-of-the-art design criteria when using higher order PSK modulations. As expected UBMSEP outperforms MMDDT criterion for all examined SNR range and outperforms the MMSE criterion for medium and high-SNR regime.
Finally, as expected the MSEP B\&B algorithm yields the same result as exhaustive search for both MSEP criteria.

\subsection{SER comparison with the state-of-the-art}

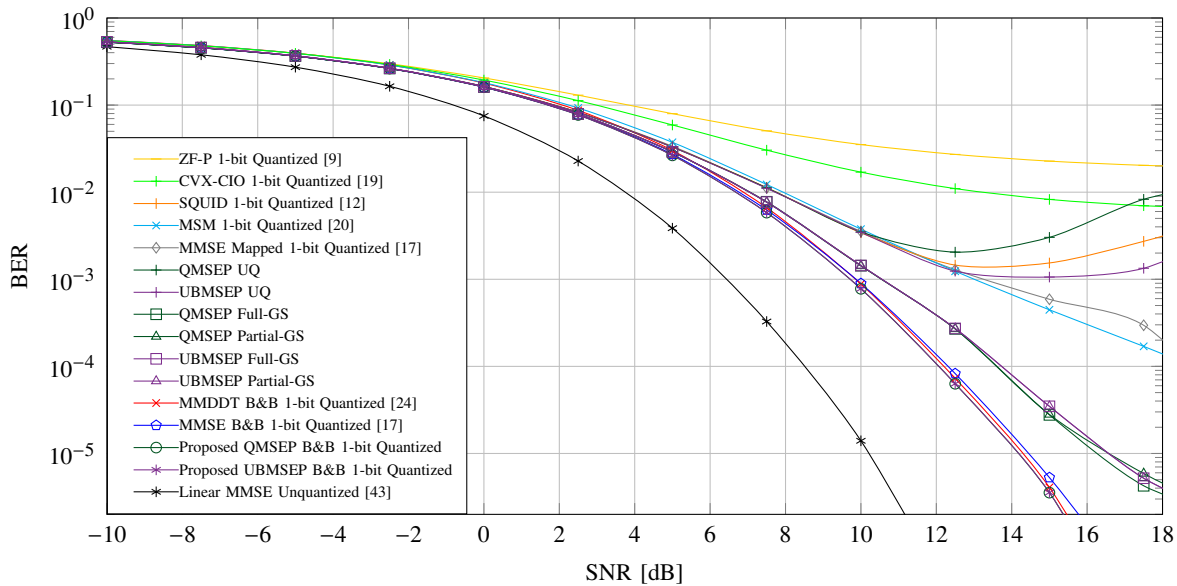
\begin{figure}[t]
\begin{center}
\input{Numerical_results/Figures/Uncoded_All_Precoders}
\caption{SER versus $\mathrm{SNR}$ for $K=3$, $M=12$, $\alpha_s=4$ and $\alpha_x=4$} 
\label{fig:Prec_comp_ser}       
\end{center}
\end{figure}

In this subsection the proposed methods are compared, in terms of SER, against the mentioned state-of-the-art techniques considering the scenario of $K=3$ users, $M=12$ BS antennas, QPSK data symbols and QPSK transmit symbols. 

As shown in Fig.~\ref{fig:Prec_comp_ser}, the proposed B\&B methods outperform all other state-of-the-art approaches in terms of SER. Note that this was expected since the proposed methods compute the optimal transmit vector using as the criterion the minimization of the SEP. Moreover, as can be seen in Fig.~\ref{fig:Prec_comp_ser} using the MSEP criteria in conjunction with PBMs based on UQ yield SER degradation for high-SNR. In this context, both Full-GS and Partial-GS algorithms mitigate the performance loss due to projection and outperform all other investigated state-of-the-art approaches.

\subsection{Complexity Analysis}

\begin{figure}[t]
\begin{center}
\input{Numerical_results/Figures/Runtime}
\caption{Runtime (ms) versus $\mathrm{SNR}$, $K=3$, $M=12$, $\alpha_s=4$ and $\alpha_x=4$ }
\label{fig:Runtime}       
\end{center}
\end{figure}
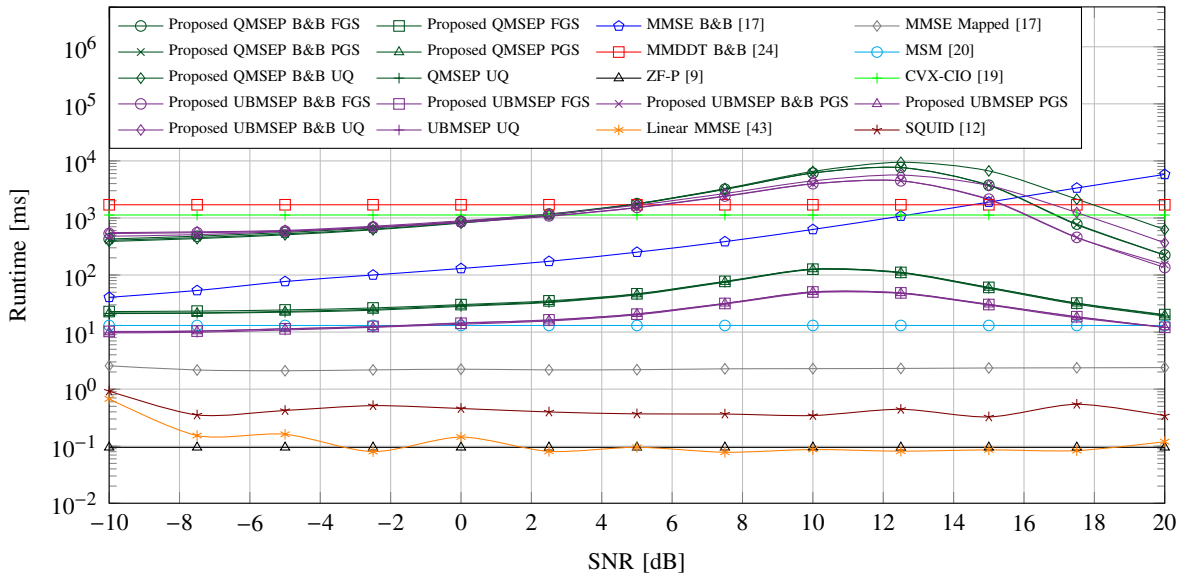
In this section, the complexity of the proposed methods is evaluated against the mentioned state-of-the art approaches considering the same scenario utilized in the previous subsection, $K=3$ users $M=12$ BS antennas and QPSK data and transmit vector symbols. The complexity of the proposed B\&B approaches is measured using the different methods for projection presented in section \ref{sec:PBM_MSEP}.
As mentioned before, the proposed methods are evaluated against the state-of-the-art approaches in terms of the runtime. For solving the optimization problems for the methods in \cite{MSM_precoder,General_MMDDT_BB,MMSE_bb} and for the proposed approaches the optimization toolbox from Matlab was utilized. The optimization problem required for the CVX-CIO approach \cite{CVX-CIO} was solved via CVX \cite{Boyd_2004}, as suggested by the authors. The results are shown in Fig.~\ref{fig:Runtime}.

As can be seen in Fig.~\ref{fig:Runtime} the proposed MSEP-GS methods have similar runtime when compared with the MSEP projection methods using UQ. Moreover, the proposed MSEP-GS approaches have significantly smaller runtime when compared with the optimal B\&B techniques which confirms the suitability of the proposed MSEP-GS methods as reasonable complexity performance tradeoff approaches. As also shown in Fig.~\ref{fig:Runtime} the proposed B\&B methods have smaller runtime than the MMDDT B\&B approach from \cite{Landau2017} for most of the SNR range analyzed and also outperforms the MMSE B\&B approach from \cite{MMSE_bb} for high SNR.

%% file: Numerical_results/Figures/Uncoded_BB_Precoders.tex
%
%
%
\usetikzlibrary{positioning,calc}
\usetikzlibrary{spy}

\definecolor{mycolor1}{rgb}{0.00000,1.00000,1.00000}%
\definecolor{mycolor2}{rgb}{1.00000,0.00000,1.00000}%

\pgfplotsset{every axis label/.append style={font=\footnotesize},
every tick label/.append style={font=\footnotesize}
}

\begin{tikzpicture}[spy using outlines={rectangle, magnification=8,connect spies}]

\begin{axis}[
name=ber,
ymode=log,
width  = 0.85\columnwidth,
height = 0.4\columnwidth,
scale only axis,
xmin  = -10,
xmax  = 20,
xlabel= {SNR  [dB]},
xmajorgrids,
ymin=5e-4,
ymax=0.61,
ylabel={SER},
ymajorgrids,
legend entries={
                },
legend style={at={(0,0.33)},anchor=south west,draw=black,fill=white,legend cell align=left,font=\tiny}
]

\addlegendimage{solid,no marks,color=black,fill=gray!20,mark=square}


\addplot+[smooth,color=r,solid, every mark/.append style={solid, fill=red!20},mark=square,
y filter/.code={\pgfmathparse{\pgfmathresult-0}\pgfmathresult}]
  table[row sep=crcr]{%
  -10.0000     0.602784700000000   \\
   -7.5000     0.551491150000000   \\
   -5.0000     0.482133050000000   \\
   -2.5000     0.396801500000000   \\
         0     0.302092650000000   \\
    2.5000     0.204317650000000   \\
    5.0000     0.120720850000000   \\
    7.5000     0.0624841000000000   \\
   10.0000     0.0287933500000000   \\
   12.5000     0.0120276500000000      \\
   15.0000     0.00474560000000000      \\
   17.5000     0.00177115000000000      \\
   20.0000     0.000553700000000000      \\
} ;

\addplot+[smooth,color=blue,solid, every mark/.append style={solid, fill=blue!20},mark=asterisk,
y filter/.code={\pgfmathparse{\pgfmathresult-0}\pgfmathresult}]
  table[row sep=crcr]{%
  -10.0000     0.581713000000000      \\
   -7.5000     0.525978950000000      \\
   -5.0000     0.454249050000000      \\
   -2.5000     0.369306800000000      \\
         0     0.279711900000000      \\
    2.5000     0.187898950000000      \\
    5.0000     0.111802300000000      \\
    7.5000     0.0589347500000000     \\
   10.0000     0.0281004500000000     \\
   12.5000     0.0123792500000000        \\
   15.0000     0.00535265000000000       \\
   17.5000     0.00217050000000000       \\
   20.0000     0.000960700000000000      \\
};

\addplot+[smooth,color=dark_green,solid, every mark/.append style={solid, fill=blue!20},mark=o,
y filter/.code={\pgfmathparse{\pgfmathresult-0}\pgfmathresult}]
  table[row sep=crcr]{%
  -10.0000     0.582056750000000    \\
   -7.5000     0.526767700000000    \\
   -5.0000     0.455000150000000    \\
   -2.5000     0.370269350000000    \\
         0     0.279507850000000    \\
    2.5000     0.187388450000000    \\
    5.0000     0.110162550000000    \\
    7.5000     0.0566899000000000    \\
   10.0000     0.0259277500000000      \\
   12.5000     0.0107689000000000      \\
   15.0000     0.00422820000000000      \\
   17.5000     0.00160975000000000      \\
   20.0000     0.000514200000000000      \\
};

\addplot+[smooth,color=dark_green,solid, every mark/.append style={solid, fill=blue!20},mark=+,
y filter/.code={\pgfmathparse{\pgfmathresult-0}\pgfmathresult}]
  table[row sep=crcr]{%
  -10.0000    0.582056750000000    \\
   -7.5000    0.526767700000000    \\
   -5.0000    0.455000150000000    \\
   -2.5000    0.370269350000000    \\
         0    0.279507850000000    \\
    2.5000    0.187388450000000    \\
    5.0000    0.110162550000000    \\
    7.5000    0.0566899000000000    \\
   10.0000    0.0259277500000000     \\
   12.5000    0.0107689000000000     \\
   15.0000    0.00422820000000000     \\
   17.5000    0.00160975000000000     \\
   20.0000    0.000514200000000000     \\
};

\addplot+[smooth,color=purple,solid, every mark/.append style={solid, fill=blue!20},mark=o,
y filter/.code={\pgfmathparse{\pgfmathresult-0}\pgfmathresult}]
  table[row sep=crcr]{%
  -10.0000     0.585653400000000   \\
   -7.5000     0.530721100000000   \\
   -5.0000     0.459232200000000   \\
   -2.5000     0.374014850000000   \\
         0     0.281457050000000   \\
    2.5000     0.188261950000000   \\
    5.0000     0.110536200000000   \\
    7.5000     0.0567985000000000   \\
   10.0000     0.0259277500000000      \\
   12.5000     0.0107092500000000      \\
   15.0000     0.00422900000000000      \\
   17.5000     0.00160980000000000      \\
   20.0000     0.000514200000000000      \\
};

\addplot+[smooth,color=purple,solid, every mark/.append style={solid, fill=blue!20},mark=+,
y filter/.code={\pgfmathparse{\pgfmathresult-0}\pgfmathresult}]
  table[row sep=crcr]{%
  -10.0000      0.585828600000000  \\
   -7.5000      0.530883400000000  \\
   -5.0000      0.459246000000000  \\
   -2.5000      0.374030850000000  \\
         0      0.281542100000000  \\
    2.5000      0.188261950000000  \\
    5.0000      0.110536200000000  \\
    7.5000      0.0567985000000000  \\
   10.0000      0.0259277500000000     \\
   12.5000      0.0107092500000000     \\
   15.0000      0.00422900000000000     \\
   17.5000      0.00160975000000000     \\
   20.0000      0.000514200000000000     \\
};

\addplot[smooth,color=red,solid,mark=square,
y filter/.code={\pgfmathparse{\pgfmathresult-0}\pgfmathresult}]
  table[row sep=crcr]{%
	1 2\\
};\label{plot1}

\addplot[smooth,color=blue,solid, mark=asterisk,
y filter/.code={\pgfmathparse{\pgfmathresult-0}\pgfmathresult}]
  table[row sep=crcr]{%
	1 2\\
};\label{plot2}

\addplot[smooth,color=dark_green,solid, mark=o,
y filter/.code={\pgfmathparse{\pgfmathresult-0}\pgfmathresult}]
  table[row sep=crcr]{%
	1 2\\
};\label{plot_QMSEP_BB}

\addplot[smooth,color=dark_green,solid, mark=+,
y filter/.code={\pgfmathparse{\pgfmathresult-0}\pgfmathresult}]
  table[row sep=crcr]{%
	1 2\\
};\label{plot_QMSEP_Ex}

\addplot[smooth,color=purple,solid, mark=o,
y filter/.code={\pgfmathparse{\pgfmathresult-0}\pgfmathresult}]
  table[row sep=crcr]{%
	1 2\\
};\label{plot_UBMSEP_BB}

\addplot[smooth,color=purple,solid, mark=+,
y filter/.code={\pgfmathparse{\pgfmathresult-0}\pgfmathresult}]
  table[row sep=crcr]{%
	1 2\\
};\label{plot_UBMSEP_Ex}

\coordinate (spypoint) at (axis cs:0,0.29);
\coordinate (spyviewer) at (axis cs:-2.5,3e-2);
\spy[width=5cm,height=2cm] on (spypoint) in node [fill=white] at (spyviewer);

\node [draw,fill=white,font=\tiny,anchor= south  west] at (axis cs: -10,5*10^-4) {
\setlength{\tabcolsep}{0.5mm}
\renewcommand{\arraystretch}{.8}
\begin{tabular}{l}

\ref{plot1}{Exhaustive Search using the MMDDT criterion} \\
\ref{plot2}{Exhaustive Search using the MMSE criterion}\\
\ref{plot_QMSEP_BB}{Proposed QMSEP B\&B}\\
\ref{plot_QMSEP_Ex}{Exhaustive Search using the QMSEP criterion}\\
\ref{plot_UBMSEP_BB}{Proposed UBMSEP B\&B}\\
\ref{plot_UBMSEP_Ex}{Exhaustive Search using the UBMSEP criterion}\\
\end{tabular}
};

\end{axis}

\end{tikzpicture}%

%% file: Numerical_results/Figures/Uncoded_BB.tex
%
%
%
\usetikzlibrary{positioning,calc}
\usetikzlibrary{spy}

\definecolor{mycolor1}{rgb}{0.00000,1.00000,1.00000}%
\definecolor{mycolor2}{rgb}{1.00000,0.00000,1.00000}%

\pgfplotsset{every axis label/.append style={font=\footnotesize},
every tick label/.append style={font=\footnotesize}
}
\begin{tikzpicture}[] 

\begin{axis}[%
name=ber,
ymode=log,
width  = 0.85\columnwidth,
height = 0.4\columnwidth,
scale only axis,
xmin  = -10,
xmax  = 20,
xlabel= {SNR  [dB]},
xmajorgrids,
ymin=1.8e-3,
ymax=1,
ylabel={SER},
ymajorgrids,
legend entries={
                },
legend style={at={(0,0.33)},anchor=south west,draw=black,fill=white,legend cell align=left,font=\tiny}
]

\addlegendimage{solid,no marks,color=black,fill=gray!20,mark=square}


\addplot+[smooth,color=red,solid, every mark/.append style={solid, fill=red!20},mark=+,
y filter/.code={\pgfmathparse{\pgfmathresult-0}\pgfmathresult}]
  table[row sep=crcr]{%
  -10.0000     0.786604750000000     \\
   -7.5000     0.750749500000000     \\
   -5.0000     0.699272750000000     \\
   -2.5000     0.635316750000000     \\
         0     0.541805000000000     \\
    2.5000     0.434782250000000     \\
    5.0000     0.314838500000000     \\
    7.5000     0.200871500000000     \\
   10.0000     0.109576500000000     \\
   12.5000     0.0489227500000000     \\
   15.0000     0.0189822500000000     \\
   17.5000     0.00656175000000000     \\
   20.0000     0.00219525000000000     \\
} ;

\addplot+[smooth,color=blue,solid, every mark/.append style={solid, fill=blue!20},mark=+,
y filter/.code={\pgfmathparse{\pgfmathresult-0}\pgfmathresult}]
  table[row sep=crcr]{%
  -10.0000    0.770213500000000    \\
   -7.5000    0.730149250000000    \\
   -5.0000    0.674250000000000    \\
   -2.5000    0.608006500000000    \\
         0    0.514601000000000    \\
    2.5000    0.411997500000000    \\
    5.0000    0.299399000000000    \\
    7.5000    0.194059500000000    \\
   10.0000    0.109002500000000    \\
   12.5000    0.0515655000000000    \\
   15.0000    0.0216840000000000    \\
   17.5000    0.00919825000000000    \\
   20.0000    0.00427225000000000    \\
};

\addplot+[smooth,color=purple,solid, every mark/.append style={solid, fill=blue!20},mark=o,
y filter/.code={\pgfmathparse{\pgfmathresult-0}\pgfmathresult}]
  table[row sep=crcr]{%
  -10.0000    0.772625000000000     \\
   -7.5000    0.732374500000000     \\
   -5.0000    0.675830000000000     \\
   -2.5000    0.607292250000000     \\
         0    0.510703000000000     \\
    2.5000    0.405264250000000     \\
    5.0000    0.290417500000000     \\
    7.5000    0.183837000000000     \\
   10.0000    0.100056500000000     \\
   12.5000    0.0447077500000000     \\
   15.0000    0.0173600000000000     \\
   17.5000    0.00600325000000000     \\
   20.0000    0.00198775000000000     \\
};

\addplot+[smooth,color=purple,solid, every mark/.append style={solid, fill=blue!20},mark=+,
y filter/.code={\pgfmathparse{\pgfmathresult-0}\pgfmathresult}]
  table[row sep=crcr]{%
  -10.0000     0.772587000000000     \\
   -7.5000     0.732337250000000     \\
   -5.0000     0.675847500000000     \\
   -2.5000     0.607297250000000     \\
         0     0.510669750000000     \\
    2.5000     0.405170000000000     \\
    5.0000     0.290405750000000     \\
    7.5000     0.183753000000000     \\
   10.0000     0.100026000000000     \\
   12.5000     0.0446615000000000     \\
   15.0000     0.0173600000000000     \\
   17.5000     0.00600325000000000     \\
   20.0000     0.00198775000000000     \\
};

\addplot[smooth,color=red,solid,mark=+,
y filter/.code={\pgfmathparse{\pgfmathresult-0}\pgfmathresult}]
  table[row sep=crcr]{%
	1 2\\
};\label{plot1}

\addplot[smooth,color=blue,solid, mark=+,
y filter/.code={\pgfmathparse{\pgfmathresult-0}\pgfmathresult}]
  table[row sep=crcr]{%
	1 2\\
};\label{plot2}

\addplot[smooth,color=purple,solid, mark=o,
y filter/.code={\pgfmathparse{\pgfmathresult-0}\pgfmathresult}]
  table[row sep=crcr]{%
	1 2\\
};\label{plot_UBMSEP_BB}

\addplot[smooth,color=purple,solid, mark=+,
y filter/.code={\pgfmathparse{\pgfmathresult-0}\pgfmathresult}]
  table[row sep=crcr]{%
	1 2\\
};\label{plot_UBMSEP_Ex}


\node [draw,fill=white,font=\tiny,anchor= south  west] at (axis cs: -10,1.8e-3) {
\setlength{\tabcolsep}{0.5mm}
\renewcommand{\arraystretch}{.8}
\begin{tabular}{l}

\ref{plot1}{Exhaustive Search using the MMDDT criterion} \\
\ref{plot2}{Exhaustive Search using the MMSE criterion}\\
\ref{plot_UBMSEP_BB}{Proposed UBMSEP B\&B}\\
\ref{plot_UBMSEP_Ex}{Exhaustive Search using the UB-MSEP criterion}\\
\end{tabular}
};

\end{axis}

\end{tikzpicture}%

%% file: Numerical_results/Figures/Uncoded_All_Precoders.tex
%
%
%
\usetikzlibrary{positioning,calc}

\definecolor{mycolor1}{rgb}{0.00000,1.00000,1.00000}%
\definecolor{mycolor2}{rgb}{1.00000,0.00000,1.00000}%

\pgfplotsset{every axis label/.append style={font=\footnotesize},
every tick label/.append style={font=\footnotesize}
}

\begin{tikzpicture}[font=\footnotesize] 

\begin{axis}[%
name=ber,
ymode=log,
width  = 0.85\columnwidth,
height = 0.4\columnwidth,
scale only axis,
xmin  = -10,
xmax  = 18,
xlabel= {SNR  [dB]},
xmajorgrids,
ymin=2e-6,
ymax=1,
ylabel={BER},
ymajorgrids,
legend entries={
                },
legend style={at={(0,0.33)},anchor=south west,draw=black,fill=white,legend cell align=left,font=\tiny}
]

\addlegendimage{solid,no marks,color=black,fill=gray!20,mark=square}


\addplot+[smooth,color=gray,solid, every mark/.append style={solid, fill=gray!20},mark=diamond,
y filter/.code={\pgfmathparse{\pgfmathresult-0}\pgfmathresult}]
table[row sep=crcr]{%
-10	    0.527494550000000         \\
-7.5    0.454275058333333         \\
-5	    0.364861950000000         \\
-2.5    0.263584983333333         \\
0	    0.163381900000000         \\
2.5	    0.0824450750000000        \\
5	    0.0336340916666667        \\
7.5	    0.0111542833333333        \\
10	    0.00354287500000000       \\
12.5    0.00128641666666667       \\
15	    0.000589366666666667          \\
17.5    0.000296475000000000          \\
20	    2.77083333333333e-05          \\
};

\addplot+[smooth,color=red,solid, every mark/.append style={solid, fill=red!20},mark=x,
y filter/.code={\pgfmathparse{\pgfmathresult-0}\pgfmathresult}]
  table[row sep=crcr]{%
-10	    0.552776358333333  \\
-7.5    0.483215708333333  \\
-5	    0.394527191666667  \\
-2.5    0.288893650000000  \\
0	    0.178867291666667  \\
2.5	    0.0869489250000000  \\
5	    0.0301083416666667  \\
7.5	    0.00663528333333333  \\
10	    0.000881466666666667  \\
12.5    7.28416666666667e-05  \\
15	    4.07500000000000e-06  \\
17.5    5.83333333333333e-08  \\
};

\addplot+[smooth,color=blue,solid, every mark/.append style={solid, fill=blue!20},mark=pentagon,
y filter/.code={\pgfmathparse{\pgfmathresult-0}\pgfmathresult}]
  table[row sep=crcr]{%
-10	    0.527471000000000         \\
-7.5    0.454196600000000         \\
-5	    0.364573141666667         \\
-2.5    0.262495866666667         \\
0	    0.160492500000000         \\
2.5	    0.0776328166666667        \\
5	    0.0271812583333333        \\
7.5	    0.00625005833333333       \\
10	    0.000894591666666667          \\
12.5    8.26666666666667e-05          \\
15	    5.29166666666667e-06          \\
17.5    2.08333333333333e-07          \\
20	    8.33333333333333e-09          \\
};

\addplot+[smooth,color=aureolin,solid, every mark/.append style={solid, fill=black!20},mark=-,
y filter/.code={\pgfmathparse{\pgfmathresult-0}\pgfmathresult}]
  table[row sep=crcr]{%
-10	   0.546802875000000  \\
-7.5   0.478175641666667  \\
-5	   0.393591616666667  \\
-2.5   0.298256000000000  \\
0	   0.205168425000000  \\
2.5	   0.130013858333333  \\
5	   0.0797388333333333  \\
7.5	   0.0506300583333333  \\
10	   0.0352139500000000  \\
12.5   0.0271405083333333  \\
15	   0.0227379166666667  \\
17.5   0.0203267333333333  \\
20	   0.0189594250000000  \\
};

\addplot+[smooth,color=cyan,solid, every mark/.append style={solid, fill=cyan!50},mark=x,
y filter/.code={\pgfmathparse{\pgfmathresult-0}\pgfmathresult}]
  table[row sep=crcr]{%
-10	   0.551066900000000      \\
-7.5   0.481328500000000      \\
-5	   0.392750933333333      \\
-2.5   0.288220391666667      \\
0	   0.180983375000000      \\
2.5	   0.0926000250000000     \\
5	   0.0373817083333333     \\
7.5	   0.0122637583333333     \\
10	   0.00376221666666667    \\
12.5   0.00123737500000000    \\
15	   0.000446058333333333   \\
17.5   0.000169791666666667   \\
20	   6.22750000000000e-05   \\
};

\addplot+[smooth,color=green,solid, every mark/.append style={solid, fill=green!50},mark=+,
y filter/.code={\pgfmathparse{\pgfmathresult-0}\pgfmathresult}]
  table[row sep=crcr]{%
-10	  0.548796175000000  \\
-7.5  0.479614716666667  \\
-5	  0.393071550000000  \\
-2.5  0.293469333333333  \\
0	  0.193913650000000  \\
2.5	  0.112524750000000  \\
5	  0.0591299833333333  \\
7.5	  0.0303856916666667  \\
10	  0.0169865416666667  \\
12.5  0.0109957500000000  \\
15	  0.00826040833333333  \\
17.5  0.00698865000000000  \\
20	  0.00638372500000000  \\
};

\addplot+[smooth,color=orange,solid, every mark/.append style={solid, fill=green!50},mark=|,
y filter/.code={\pgfmathparse{\pgfmathresult-0}\pgfmathresult}]
  table[row sep=crcr]{%
-10	   0.529371058333333  \\
-7.5   0.456290441666667  \\
-5	   0.366414500000000  \\
-2.5   0.264253833333333  \\
0	   0.163483816666667  \\
2.5	   0.0829327916666667  \\
5	   0.0330774333333333  \\
7.5	   0.0112671083333333  \\
10	   0.00342175833333333  \\
12.5   0.00145604166666667  \\
15	   0.00153996666666667  \\
17.5   0.00272738333333333  \\
20	   0.00563230000000000  \\
};

 \addplot+[smooth,color=black,solid, every mark/.append style={solid, fill=blue!20},mark=asterisk,
 y filter/.code={\pgfmathparse{\pgfmathresult-0}\pgfmathresult}]
   table[row sep=crcr]{%
-10	    0.468178033333333  \\
-7.5    0.378037375000000  \\
-5	    0.272607916666667  \\
-2.5    0.164424425000000  \\
0	    0.0751228000000000  \\
2.5	    0.0227874083333333  \\
5	    0.00386170000000000  \\
7.5	    0.000327475000000000  \\
10	    1.39833333333333e-05  \\
12.5    1.83333333333333e-07  \\
 };             

 \addplot+[smooth,color=dark_green ,solid, every mark/.append style={solid, fill=cyan!20},mark=o,
 y filter/.code={\pgfmathparse{\pgfmathresult-0}\pgfmathresult}]
   table[row sep=crcr]{%
-10	   0.528216341666667   \\
-7.5   0.455165133333333   \\
-5	   0.365330800000000   \\
-2.5   0.262942225000000   \\
0	   0.160348533333333   \\
2.5	   0.0770420166666667   \\
5	   0.0264610250000000   \\
7.5	   0.00582552500000000   \\
10	   0.000777350000000000   \\
12.5   6.31583333333333e-05   \\
15	   3.55000000000000e-06   \\
17.5   5.00000000000000e-08   \\
 };

 \addplot+[smooth,color=dark_green ,solid, every mark/.append style={solid, fill=cyan!20},mark=+,
 y filter/.code={\pgfmathparse{\pgfmathresult-0}\pgfmathresult}]
   table[row sep=crcr]{%
-10	   0.528201908333333   \\
-7.5   0.455186950000000   \\
-5	   0.365589541666667   \\
-2.5   0.264152208333333   \\
0	   0.163332400000000   \\
2.5	   0.0824393166666667   \\
5	   0.0329374250000000   \\
7.5	   0.0112374333333333   \\
10	   0.00349615833333333   \\
12.5   0.00204996666666667   \\
15	   0.00302932500000000   \\
17.5   0.00828536666666667   \\
20	   0.0134229166666667   \\            
 };

 \addplot+[smooth,color=dark_green ,solid, every mark/.append style={solid, fill=cyan!20},mark=square,
 y filter/.code={\pgfmathparse{\pgfmathresult-0}\pgfmathresult}]
   table[row sep=crcr]{%
-10	    0.528205508333333  \\
-7.5    0.455187700000000  \\
-5	    0.365496333333333  \\
-2.5    0.263302333333333  \\
0	    0.161353283333333  \\
2.5	    0.0790331166666667  \\
5	    0.0290511000000000  \\
7.5	    0.00776655000000000  \\
10	    0.00143296666666667  \\
12.5    0.000269650000000000  \\
15	    2.77416666666667e-05  \\
17.5    4.25833333333333e-06  \\
20	    1.88333333333333e-06  \\            
 };

 \addplot+[smooth,color=dark_green ,solid, every mark/.append style={solid, fill=cyan!20},mark=triangle,
 y filter/.code={\pgfmathparse{\pgfmathresult-0}\pgfmathresult}]
   table[row sep=crcr]{%
-10	    0.528204400000000  \\
-7.5    0.455190858333333  \\
-5	    0.365473358333333  \\
-2.5    0.263264808333333  \\
0	    0.161412208333333  \\
2.5	    0.0791413750000000  \\
5	    0.0289364583333333  \\
7.5	    0.00766581666666667  \\
10	    0.00142230000000000  \\
12.5    0.000269083333333333  \\
15	    2.80333333333333e-05  \\
17.5    5.89166666666667e-06  \\
20	    1.80833333333333e-06  \\            
 };

 \addplot+[smooth,color=purple ,solid, every mark/.append style={solid, fill=cyan!20},mark=+,
 y filter/.code={\pgfmathparse{\pgfmathresult-0}\pgfmathresult}]
   table[row sep=crcr]{%
-10	   0.531843025000000      \\
-7.5   0.458614333333333      \\
-5	   0.368190158333333      \\
-2.5   0.265440941666667      \\
0	   0.163729333333333      \\
2.5	   0.0824711833333333     \\
5	   0.0329597083333333     \\
7.5	   0.0110851083333333     \\
10	   0.00346095000000000    \\
12.5   0.00123960000000000        \\
15	   0.00106065000000000        \\
17.5   0.00133737500000000        \\
20	   0.00402359166666667        \\            
 };

 \addplot+[smooth,color=purple ,solid, every mark/.append style={solid, fill=cyan!20},mark=triangle,
 y filter/.code={\pgfmathparse{\pgfmathresult-0}\pgfmathresult}]
   table[row sep=crcr]{%
-10	    0.531996400000000     \\
-7.5    0.458740250000000     \\
-5	    0.367994366666667     \\
-2.5    0.264740666666667     \\
0	    0.161870658333333     \\
2.5	    0.0791525500000000    \\
5	    0.0290827583333333    \\
7.5	    0.00769250000000000   \\
10	    0.00143984166666667   \\
12.5    0.000272725000000000      \\
15	    3.47833333333333e-05      \\
17.5    5.20000000000000e-06      \\
20	    1.70833333333333e-06      \\            
 };

 \addplot+[smooth,color=purple ,solid, every mark/.append style={solid, fill=cyan!20},mark=square,
 y filter/.code={\pgfmathparse{\pgfmathresult-0}\pgfmathresult}]
   table[row sep=crcr]{%
-10	    0.531996400000000     \\
-7.5    0.458740250000000     \\
-5	    0.367994366666667     \\
-2.5    0.264740666666667     \\
0	    0.161870658333333     \\
2.5	    0.0791525500000000    \\
5	    0.0290827583333333    \\
7.5	    0.00769250000000000   \\
10	    0.00143984166666667   \\
12.5    0.000272725000000000      \\
15	    3.47833333333333e-05      \\
17.5    5.20000000000000e-06      \\
20	    1.70833333333333e-06      \\            
 };

 \addplot+[smooth,color=purple ,solid, every mark/.append style={solid, fill=cyan!20},mark=asterisk,
 y filter/.code={\pgfmathparse{\pgfmathresult-0}\pgfmathresult}]
   table[row sep=crcr]{%
-10	  0.532008166666667     \\
-7.5  0.458724533333333     \\
-5	  0.367981266666667     \\
-2.5  0.264356450000000     \\
0	  0.160821141666667     \\
2.5	  0.0771190333333333         \\
5	  0.0264821500000000         \\
7.5	  0.00583020000000000         \\
10	  0.000777441666666667         \\
12.5  6.31833333333333e-05         \\
15	  3.55833333333333e-06         \\
17.5  5.00000000000000e-08         \\
20	  8.33333333333333e-09         \\            
 };

\addplot[smooth,color=gray,solid,mark=diamond,
y filter/.code={\pgfmathparse{\pgfmathresult-0}\pgfmathresult}]
  table[row sep=crcr]{%
	1 2\\
};\label{mmse_mapped_QQ}

\addplot[smooth,color=red,solid,mark=x,
y filter/.code={\pgfmathparse{\pgfmathresult-0}\pgfmathresult}]
  table[row sep=crcr]{%
	1 2\\
};\label{mmddt_bb_QQ}

\addplot[smooth,color=blue,solid,mark=pentagon,
y filter/.code={\pgfmathparse{\pgfmathresult-0}\pgfmathresult}]
  table[row sep=crcr]{%
	1 2\\
};\label{mmse_bb_QQ}

\addplot[smooth,color=aureolin,solid,mark=-,
y filter/.code={\pgfmathparse{\pgfmathresult-0}\pgfmathresult}]
  table[row sep=crcr]{%
	1 2\\
};\label{zfp_QQ}

\addplot[smooth,color=cyan,solid,mark=x,
y filter/.code={\pgfmathparse{\pgfmathresult-0}\pgfmathresult}]
  table[row sep=crcr]{%
	1 2\\
};\label{msm_QQ}

\addplot[smooth,color=green,solid,mark=+,
y filter/.code={\pgfmathparse{\pgfmathresult-0}\pgfmathresult}]
  table[row sep=crcr]{%
	1 2\\
};\label{cvx_cio_QQ}

\addplot[smooth,color=orange,solid,mark=|,
y filter/.code={\pgfmathparse{\pgfmathresult-0}\pgfmathresult}]
  table[row sep=crcr]{%
	1 2\\
};\label{squid_QQ}

\addplot[smooth,color=black,solid,mark=asterisk,
y filter/.code={\pgfmathparse{\pgfmathresult-0}\pgfmathresult}]
  table[row sep=crcr]{%
	1 2\\
};\label{Linear_mmse_QQ}

\addplot[smooth,color=dark_green,solid,mark=o,
y filter/.code={\pgfmathparse{\pgfmathresult-0}\pgfmathresult}]
  table[row sep=crcr]{%
	1 2\\
};\label{qmsep_bb_QQ}

\addplot[smooth,color=dark_green,solid,mark=+,
y filter/.code={\pgfmathparse{\pgfmathresult-0}\pgfmathresult}]
  table[row sep=crcr]{%
	1 2\\
};\label{qmsep_mapped_QQ}

\addplot[smooth,color=dark_green,solid,mark=triangle,
y filter/.code={\pgfmathparse{\pgfmathresult-0}\pgfmathresult}]
  table[row sep=crcr]{%
	1 2\\
};\label{qmsep_fgs_QQ}

\addplot[smooth,color=dark_green,solid,mark=square,
y filter/.code={\pgfmathparse{\pgfmathresult-0}\pgfmathresult}]
  table[row sep=crcr]{%
	1 2\\
};\label{qmsep_pgs_QQ}

\addplot[smooth,color=purple,solid,mark=+,
y filter/.code={\pgfmathparse{\pgfmathresult-0}\pgfmathresult}]
  table[row sep=crcr]{%
	1 2\\
};\label{ubmsep_mapped_QQ}

\addplot[smooth,color=purple,solid,mark=triangle,
y filter/.code={\pgfmathparse{\pgfmathresult-0}\pgfmathresult}]
  table[row sep=crcr]{%
	1 2\\
};\label{ubmsep_pgs_QQ}

\addplot[smooth,color=purple,solid,mark=square,
y filter/.code={\pgfmathparse{\pgfmathresult-0}\pgfmathresult}]
  table[row sep=crcr]{%
	1 2\\
};\label{ubmsep_fgs_QQ}

\addplot[smooth,color=purple,solid,mark=asterisk,
y filter/.code={\pgfmathparse{\pgfmathresult-0}\pgfmathresult}]
  table[row sep=crcr]{%
	1 2\\
};\label{ubmsep_bb_QQ}

\node [draw,fill=white,font=\tiny,anchor= south  west] at (axis cs: -10,2e-6) {
\setlength{\tabcolsep}{0.5mm}
\renewcommand{\arraystretch}{.8}
\begin{tabular}{l}

\ref{zfp_QQ}{ZF-P 1-bit Quantized \cite{ZF-Precoding}}\\
\ref{cvx_cio_QQ}{CVX-CIO 1-bit Quantized \cite{CVX-CIO}}\\
\ref{squid_QQ}{SQUID 1-bit Quantized} \cite{Squid_precoder}\\
\ref{msm_QQ}{MSM 1-bit Quantized \cite{MSM_precoder}}\\
\ref{mmse_mapped_QQ}{MMSE Mapped 1-bit Quantized}\cite{MMSE_bb}\\
\ref{qmsep_mapped_QQ}{QMSEP UQ}\\
\ref{ubmsep_mapped_QQ}{UBMSEP UQ}\\
\ref{qmsep_pgs_QQ}{QMSEP Full-GS}\\
\ref{qmsep_fgs_QQ}{QMSEP Partial-GS}\\
\ref{ubmsep_fgs_QQ}{UBMSEP Full-GS}\\
\ref{ubmsep_pgs_QQ}{UBMSEP Partial-GS}\\
\ref{mmddt_bb_QQ}{MMDDT B\&B 1-bit Quantized} \cite{Landau2017}\\
\ref{mmse_bb_QQ}{MMSE B\&B 1-bit Quantized} \cite{MMSE_bb}\\
\ref{qmsep_bb_QQ}{Proposed QMSEP B\&B 1-bit Quantized}\\
\ref{ubmsep_bb_QQ}{Proposed UBMSEP B\&B  1-bit Quantized}\\
\ref{Linear_mmse_QQ}{Linear MMSE Unquantized \cite{M_Joham_ZF}}
\end{tabular}
};

\end{axis}

\end{tikzpicture}%

%% file: Numerical_results/Figures/Runtime.tex
%
%
%
\usetikzlibrary{positioning,calc}

\definecolor{mycolor1}{rgb}{0.00000,1.00000,1.00000}%
\definecolor{mycolor2}{rgb}{1.00000,0.00000,1.00000}%

\pgfplotsset{every axis label/.append style={font=\footnotesize},
every tick label/.append style={font=\footnotesize}
}

\begin{tikzpicture}[font=\footnotesize] 

\begin{axis}[%
name=ber,
ymode=log,
width  = 0.85\columnwidth,
height = 0.4\columnwidth,
scale only axis,
xmin  = -10,
xmax  = 20,
xlabel= {SNR  [dB]},
xmajorgrids,
ymin=10^-2,
ymax=5*10^6,
ylabel={Runtime [ms]},
ymajorgrids,
legend entries={,
Proposed QMSEP B\&B FGS,
Proposed QMSEP FGS,
MMSE B\&B \cite{MMSE_bb},
MMSE Mapped \cite{MMSE_bb},
Proposed QMSEP B\&B PGS,
Proposed QMSEP PGS,
MMDDT B\&B \cite{Landau2017},
MSM \cite{MSM_precoder},
Proposed QMSEP B\&B UQ,
QMSEP UQ,
ZF-P  \cite{ZF-Precoding},
CVX-CIO  \cite{CVX-CIO},
Proposed UBMSEP B\&B FGS,
Proposed UBMSEP FGS,
Proposed UBMSEP B\&B PGS,
Proposed UBMSEP PGS,
Proposed UBMSEP B\&B UQ,
UBMSEP UQ,
Linear MMSE  \cite{M_Joham_ZF},
SQUID \cite{Squid_precoder},
},
legend columns=4,
legend style={at={(0,1)},anchor=north west,draw=black,fill=white,legend cell align=left,font=\tiny,}
]

\addlegendimage{solid,no marks,color=black,fill=gray!20,mark=square}


\definecolor{dark_red}{rgb}{0.5,0,0}

\addplot+[smooth,color=dark_green,solid, every mark/.append style={solid, fill=gray!20},mark=o,
y filter/.code={\pgfmathparse{\pgfmathresult-0}\pgfmathresult}]
table[row sep=crcr]{%
  -10.0000   428.263397062500   \\
   -7.5000   479.675227000000   \\
   -5.0000   556.882853687500   \\
   -2.5000   670.936319250000   \\
         0   866.035056000000   \\
    2.5000   1172.37884975000   \\
    5.0000   1782.25195543750   \\
    7.5000   3195.33934931250   \\
   10.0000   6119.02023987500   \\
   12.5000   7671.98460837500   \\
   15.0000   3741.94042318750   \\
   17.5000   780.603499937500   \\
   20.0000   224.338568625000   \\
};

\addplot+[smooth,color=dark_green,solid, every mark/.append style={solid, fill=gray!20},mark=square,
y filter/.code={\pgfmathparse{\pgfmathresult-0}\pgfmathresult}]
table[row sep=crcr]{%
  -10.0000       22.8976   \\
   -7.5000       23.3071   \\
   -5.0000       24.4303   \\
   -2.5000       26.4571   \\
         0       30.3081   \\
    2.5000       35.2307   \\
    5.0000       46.8669   \\
    7.5000       77.3707   \\
   10.0000      126.5807   \\
   12.5000      111.1619   \\
   15.0000       60.7070   \\
   17.5000       32.2003   \\
   20.0000       20.2551   \\
};

\addplot+[smooth,color=blue,solid, every mark/.append style={solid, fill=blue!20},mark=pentagon,
y filter/.code={\pgfmathparse{\pgfmathresult-0}\pgfmathresult}]
  table[row sep=crcr]{%
  -10.0000  40.6476743125000    \\
   -7.5000  53.5347860625000    \\
   -5.0000  77.0100035625000    \\
   -2.5000  100.120095000000    \\
         0  130.040655500000    \\
    2.5000  173.661585812500    \\
    5.0000  251.297054750000    \\
    7.5000  384.874225375000    \\
   10.0000  628.630284812500    \\
   12.5000  1076.69590193750    \\
   15.0000  1896.05094137500    \\
   17.5000  3348.29829450000    \\
   20.0000  5797.60126468750    \\
};

\addplot+[smooth,color=gray,solid, every mark/.append style={solid, fill=gray!20},mark=diamond,
y filter/.code={\pgfmathparse{\pgfmathresult-0}\pgfmathresult}]
table[row sep=crcr]{%
   -10.0000  2.56778450000000   \\
   -7.5000   2.16250887500000   \\
   -5.0000   2.09911862500000   \\
   -2.5000   2.17399737500000   \\
         0   2.23610287500000   \\
    2.5000   2.17298350000000   \\
    5.0000   2.19199006250000   \\
    7.5000   2.26941062500000   \\
   10.0000   2.28655975000000   \\
   12.5000   2.30509350000000   \\
   15.0000   2.34875562500000   \\
   17.5000   2.36430525000000   \\
   20.0000   2.38621718750000       \\
};

\addplot+[smooth,color=dark_green,solid, every mark/.append style={solid, fill=blue!20},mark=x,
y filter/.code={\pgfmathparse{\pgfmathresult-0}\pgfmathresult}]
  table[row sep=crcr]{%
  -10.0000   406.408197750000   \\
   -7.5000   451.976890812500   \\
   -5.0000   525.792595187500   \\
   -2.5000   636.393318562500   \\
         0   827.617706375000   \\
    2.5000   1136.70340618750   \\
    5.0000   1752.45701068750   \\
    7.5000   3162.69718850000   \\
   10.0000   6133.93607662500   \\
   12.5000   7607.68796025000   \\
   15.0000   3718.01768187500   \\
   17.5000   759.259667562500   \\
   20.0000   222.973417687500   \\
};

\addplot+[smooth,color=dark_green,solid, every mark/.append style={solid, fill=blue!20},mark=triangle,
y filter/.code={\pgfmathparse{\pgfmathresult-0}\pgfmathresult}]
  table[row sep=crcr]{%
  -10.0000       21.7266   \\
   -7.5000       21.6950   \\
   -5.0000       23.0494   \\
   -2.5000       25.0614   \\
         0       29.1839   \\
    2.5000       33.9422   \\
    5.0000       46.1659   \\
    7.5000       76.3434   \\
   10.0000      127.0245   \\
   12.5000      110.9481   \\
   15.0000       60.2523   \\
   17.5000       31.8885   \\
   20.0000       19.7358   \\
};

\addplot+[smooth,color=red,solid, every mark/.append style={solid, fill=red!20},mark=square,
y filter/.code={\pgfmathparse{\pgfmathresult-0}\pgfmathresult}]
  table[row sep=crcr]{%
  -10.0000   1702.20248468750   \\
   -7.5000   1702.20248468750   \\
   -5.0000   1702.20248468750   \\
   -2.5000   1702.20248468750   \\
         0   1702.20248468750   \\
    2.5000   1702.20248468750   \\
    5.0000   1702.20248468750   \\
    7.5000   1702.20248468750   \\
   10.0000   1702.20248468750   \\
   12.5000   1702.20248468750   \\
   15.0000   1702.20248468750   \\
   17.5000   1702.20248468750   \\
   20.0000   1702.20248468750   \\
};

\addplot+[smooth,color=cyan,solid, every mark/.append style={solid, fill=cyan!50},mark=o,
y filter/.code={\pgfmathparse{\pgfmathresult-0}\pgfmathresult}]
  table[row sep=crcr]{%
  -10.0000    13.0510000000000   \\
   -7.5000    13.0510000000000   \\
   -5.0000    13.0510000000000   \\
   -2.5000    13.0510000000000   \\
         0    13.0510000000000   \\
    2.5000    13.0510000000000   \\
    5.0000    13.0510000000000   \\
    7.5000    13.0510000000000   \\
   10.0000    13.0510000000000   \\
   12.5000    13.0510000000000   \\
   15.0000    13.0510000000000   \\
   17.5000    13.0510000000000   \\
   20.0000    13.0510000000000   \\
};

\addplot+[smooth,color=dark_green,solid, every mark/.append style={solid, fill=cyan!50},mark=diamond,
y filter/.code={\pgfmathparse{\pgfmathresult-0}\pgfmathresult}]
  table[row sep=crcr]{%
  -10.0000  387.423980187500   \\
   -7.5000  434.705161375000   \\
   -5.0000  505.777655000000   \\
   -2.5000  618.762900250000   \\
         0  811.891965687500   \\
    2.5000  1133.97908812500   \\
    5.0000  1775.99305525000   \\
    7.5000  3259.54725368750   \\
   10.0000  6527.98897531250   \\
   12.5000  9457.68978112500   \\
   15.0000  6673.83693187500   \\
   17.5000  2133.95768325000   \\
   20.0000  626.438428812500   \\
};

\addplot+[smooth,color=dark_green,solid, every mark/.append style={solid, fill=cyan!50},mark=+,
y filter/.code={\pgfmathparse{\pgfmathresult-0}\pgfmathresult}]
  table[row sep=crcr]{%
  -10.0000       21.1136    \\
   -7.5000       21.3408    \\
   -5.0000       22.3164    \\
   -2.5000       24.1100    \\
         0       28.2191    \\
    2.5000       33.0543    \\
    5.0000       44.9004    \\
    7.5000       74.8475    \\
   10.0000      124.5362    \\
   12.5000      108.5368    \\
   15.0000       58.6075    \\
   17.5000       30.7486    \\
   20.0000       19.0828    \\
};

\addplot+[smooth,color=black,solid, every mark/.append style={solid, fill=black!20},mark=triangle,
y filter/.code={\pgfmathparse{\pgfmathresult-0}\pgfmathresult}]
  table[row sep=crcr]{%
  -10.0000  0.0953312500000000    \\
   -7.5000  0.0953312500000000    \\
   -5.0000  0.0953312500000000    \\
   -2.5000  0.0953312500000000    \\
         0  0.0953312500000000    \\
    2.5000  0.0953312500000000    \\
    5.0000  0.0953312500000000    \\
    7.5000  0.0953312500000000    \\
   10.0000  0.0953312500000000    \\
   12.5000  0.0953312500000000    \\
   15.0000  0.0953312500000000    \\
   17.5000  0.0953312500000000    \\
   20.0000  0.0953312500000000    \\
};

\addplot+[smooth,color=green,solid, every mark/.append style={solid, fill=green!50},mark=+,
y filter/.code={\pgfmathparse{\pgfmathresult-0}\pgfmathresult}]
  table[row sep=crcr]{%
  -10.0000   1126.30739918750   \\
   -7.5000   1126.30739918750   \\
   -5.0000   1126.30739918750   \\
   -2.5000   1126.30739918750   \\
         0   1126.30739918750   \\
    2.5000   1126.30739918750   \\
    5.0000   1126.30739918750   \\
    7.5000   1126.30739918750   \\
   10.0000   1126.30739918750   \\
   12.5000   1126.30739918750   \\
   15.0000   1126.30739918750   \\
   17.5000   1126.30739918750   \\
   20.0000   1126.30739918750   \\
};

\addplot+[smooth,color=purple,solid, every mark/.append style={solid, fill=cyan!50},mark=o,
y filter/.code={\pgfmathparse{\pgfmathresult-0}\pgfmathresult}]
  table[row sep=crcr]{%
  -10.0000  533.404860562500   \\
   -7.5000  551.738609562500   \\
   -5.0000  588.587107125000   \\
   -2.5000  692.778364187500   \\
         0  850.648698562500   \\
    2.5000  1087.71772762500   \\
    5.0000  1528.38735937500   \\
    7.5000  2422.36671212500   \\
   10.0000  3980.02834137500   \\
   12.5000  4477.80472525000   \\
   15.0000  2131.75420206250   \\
   17.5000  457.398001062500   \\
   20.0000  135.511333125000   \\
};

\addplot+[smooth,color=purple,solid, every mark/.append style={solid, fill=cyan!50},mark=square,
y filter/.code={\pgfmathparse{\pgfmathresult-0}\pgfmathresult}]
  table[row sep=crcr]{%
  -10.0000    10.2708  \\
   -7.5000    10.4077  \\
   -5.0000    11.4003  \\
   -2.5000    12.5618  \\
         0    14.2804  \\
    2.5000    16.3683  \\
    5.0000    20.6229  \\
    7.5000    31.7444  \\
   10.0000    50.4764  \\
   12.5000    48.6398  \\
   15.0000    30.4071  \\
   17.5000    18.6172  \\
   20.0000    12.1038  \\
};

\addplot+[smooth,color=purple,solid, every mark/.append style={solid, fill=cyan!50},mark=x,
y filter/.code={\pgfmathparse{\pgfmathresult-0}\pgfmathresult}]
  table[row sep=crcr]{%
  -10.0000   478.407065125000   \\
   -7.5000   509.445987875000   \\
   -5.0000   579.441466562500   \\
   -2.5000   682.018854125000   \\
         0   831.512145250000   \\
    2.5000   1067.36005100000   \\
    5.0000   1509.90610550000   \\
    7.5000   2384.49866450000   \\
   10.0000   3949.98523093750   \\
   12.5000   4470.41565312500   \\
   15.0000   2083.31098656250   \\
   17.5000   453.209902062500   \\
   20.0000   152.640059062500   \\
};

\addplot+[smooth,color=purple,solid, every mark/.append style={solid, fill=cyan!50},mark=triangle,
y filter/.code={\pgfmathparse{\pgfmathresult-0}\pgfmathresult}]
  table[row sep=crcr]{%
  -10.0000      9.9693    \\
   -7.5000     10.3845    \\
   -5.0000     11.4682    \\
   -2.5000     12.2487    \\
         0     14.5272    \\
    2.5000     16.0557    \\
    5.0000     20.8758    \\
    7.5000     31.8166    \\
   10.0000     50.7658    \\
   12.5000     48.4589    \\
   15.0000     30.6727    \\
   17.5000     18.3945    \\
   20.0000     12.1965    \\
};

\addplot+[smooth,color=purple,solid, every mark/.append style={solid, fill=cyan!50},mark=diamond,
y filter/.code={\pgfmathparse{\pgfmathresult-0}\pgfmathresult}]
  table[row sep=crcr]{%
  -10.0000   552.537084437500  \\
   -7.5000   571.253351125000  \\
   -5.0000   604.394235000000  \\
   -2.5000   715.960884000000  \\
         0   892.634806062501  \\
    2.5000   1165.10102681250  \\
    5.0000   1675.55830043750  \\
    7.5000   2677.94061650000  \\
   10.0000   4463.21569250000  \\
   12.5000   5649.26605825000  \\
   15.0000   3696.48310643750  \\
   17.5000   1241.76189168750  \\
   20.0000   366.961949625000  \\
};

\addplot+[smooth,color=purple,solid, every mark/.append style={solid, fill=cyan!50},mark=+,
y filter/.code={\pgfmathparse{\pgfmathresult-0}\pgfmathresult}]
  table[row sep=crcr]{%
  -10.0000       9.6888   \\
   -7.5000       9.9124   \\
   -5.0000      10.8815   \\
   -2.5000      12.0923   \\
         0      13.7372   \\
    2.5000      15.6223   \\
    5.0000      20.0514   \\
    7.5000      30.9971   \\
   10.0000      49.0775   \\
   12.5000      47.4466   \\
   15.0000      29.7752   \\
   17.5000      17.7508   \\
   20.0000      12.0935   \\
};

 \addplot+[smooth,color=orange,solid, every mark/.append style={solid, fill=blue!20},mark=asterisk,
 y filter/.code={\pgfmathparse{\pgfmathresult-0}\pgfmathresult}]
   table[row sep=crcr]{%
  -10.0000   0.675998062500000 \\
   -7.5000   0.153779375000000 \\
   -5.0000   0.161776187500000 \\
   -2.5000   0.0803267500000000 \\
         0   0.144505750000000 \\
    2.5000   0.0811833125000000 \\
    5.0000   0.0956558750000000 \\
    7.5000   0.0783006875000000 \\
   10.0000   0.0873920625000000 \\
   12.5000   0.0817571875000000 \\
   15.0000   0.0861793750000000 \\
   17.5000   0.0837011875000000 \\
   20.0000   0.119762812500000 \\
 };             

\addplot+[smooth,color=dark_red,solid, every mark/.append style={solid, fill=gray!20},mark=star,
y filter/.code={\pgfmathparse{\pgfmathresult-0}\pgfmathresult}]
table[row sep=crcr]{%
  -10.0000       0.9260   \\
   -7.5000       0.3529   \\
   -5.0000       0.4229   \\
   -2.5000       0.5137   \\
         0       0.4577   \\
    2.5000       0.3980   \\
    5.0000       0.3684   \\
    7.5000       0.3662   \\
   10.0000       0.3449   \\
   12.5000       0.4431   \\
   15.0000       0.3259   \\
   17.5000       0.5426   \\
   20.0000       0.3418   \\
};

\end{axis}

\end{tikzpicture}%

%% file: 03_Conclusion.tex
\section{Conclusions}
\label{sec:conclusion}

In this study we propose a novel precoding criterion based on the direct minimization of the SEP for quantized signals with constant envelope and PSK modulation. Unlike the existing MSEP designs \cite{Mingjie_framework,Sohrabi2018,Mingjie_icassp2018,Mingjie_SSP2018,Mingjie_globalsip2018,mezghani2020massive} the proposed approaches allow for any PSK modulation order and phase quantization with any number of bits.
Using the proposed criterion and the one from \cite{mezghani2020massive} this study develops optimal precoding techniques based on a sophisticated B\&B algorithm and practical precoding methods based on GS.

Numerical results show that the proposed optimal approaches outperforms the existing methods, in terms of SER for all investigated SNR range. Numerical results also indicate that the proposed optimal methods have less computational complexity for some scenarios when compared with other B\&B approaches. The proposed GS algorithms also outperform all other state-of-the-art suboptimal methods in terms of SER with decreased complexity when compared with the optimal approaches. As such, reasonable complexity performance tradeoffs can be achieved using the proposed GS techniques.

%% file: Appendixes/Appendix_Precoder.tex
\input{Appendixes/UB_Msep_Hessian}

%% file: Appendixes/UB_Msep_Hessian.tex
\subsection{Convexity analysis of the UBMSEP objective}
\label{app:Ub_msep}

This appendix derives the conditions for convexity of the UBMSEP objective function. In this analysis it is considered the real-valued UBMSEP formulation described in \eqref{eq:rv_ubmsep_prob}. With this, the UBSMEP objective can be cast as 
\begin{align}
    &g(  \boldsymbol{x}_{\text{r}})=  -\sum_{k=1}^K \log \pc{ \text{erf}\pc{ \pc{\boldsymbol{h}_{\text{R},\theta,k}^{s^*} -\boldsymbol{h}_{\text{I},\theta,k}^{s^*}} \boldsymbol{x}_{\text{r}} }+
    \text{erf}\pc{ \pc{\boldsymbol{h}_{\text{R},\theta,k}^{s^*} +\boldsymbol{h}_{\text{I},\theta,k}^{s^*}} \boldsymbol{x}_{\text{r} } }} .
\end{align}
To simplify the notation the vectors $\boldsymbol{h}_{1,k}$ and $\boldsymbol{h}_{2,k}$ are introduced as  $\boldsymbol{h}_{1,k}=\boldsymbol{h}_{\text{R},\theta,k}^{s^*} -\boldsymbol{h}_{\text{I},\theta,k}^{s^*}$ and $\boldsymbol{h}_{2,k}=\boldsymbol{h}_{\text{R},\theta,k}^{s^*} +\boldsymbol{h}_{\text{I},\theta,k}^{s^*}$. With this, $g(  \boldsymbol{x}_{\text{r}})$ is rewritten as
\begin{align}
    &g(  \boldsymbol{x}_{\text{r}})=  -\sum_{k=1}^K \log \pc{ \text{erf}\pc{ \boldsymbol{h}_{1,k} \boldsymbol{x}_{\text{r}} }+
    \text{erf}\pc{ \boldsymbol{h}_{2,k} \boldsymbol{x}_{\text{r} } }} 
\end{align}
Convexity can be proven by evaluating the conditions in which the Hessian is positive semi-definite (PSD). To this end the in what follows the Hessian is calculated. Taking the derivative of $g(  \boldsymbol{x}_{\text{r}})$ with respect to $\boldsymbol{x}_{\text{r}}$ 
\begin{align}
    \frac{ \partial g(  \boldsymbol{x}_{\text{r}})}{ \partial  \boldsymbol{x}_{\text{r}}}&=-\sum_{k=1}^K \frac{ \partial }{ \partial  \boldsymbol{x}_{\text{r}}}\ \log \pc{ \text{erf}\pc{ \boldsymbol{h}_{1,k} \boldsymbol{x}_{\text{r}} }+ \text{erf}\pc{ \boldsymbol{h}_{2,k} \boldsymbol{x}_{\text{r} } }} 
    \\
    &=-\sum_{k=1}^K\frac{
    \frac{ \partial }{ \partial  \boldsymbol{x}_{\text{r}}}\pc{ \text{erf}\pc{ \boldsymbol{h}_{1,k} \boldsymbol{x}_{\text{r}} }+
    \text{erf}\pc{\boldsymbol{h}_{2,k} \boldsymbol{x}_{\text{r}} }}
    }
    {\text{erf}\pc{ \boldsymbol{h}_{1,k} \boldsymbol{x}_{\text{r}} }+
    \text{erf}\pc{\boldsymbol{h}_{2,k} \boldsymbol{x}_{\text{r}} }
    }
    \\
    &=-\sum_{k=1}^K\frac{\frac{2}{\sqrt{\pi}} \ e^{{-\pc{ \boldsymbol{h}_{1,k} \boldsymbol{x}_{\text{r}}}^2}} \boldsymbol{h}_{1,k}+
    \frac{2}{\sqrt{\pi}} \ e^{{-\pc{ \boldsymbol{h}_{2,k} \boldsymbol{x}_{\text{r}}}^2}} \boldsymbol{h}_{2,k}
    }{\text{erf}\pc{ \boldsymbol{h}_{1,k} \boldsymbol{x}_{\text{r}} }+
    \text{erf}\pc{\boldsymbol{h}_{2,k} \boldsymbol{x}_{\text{r}} }}.
\end{align}
Note that $\frac{ \partial g(  \boldsymbol{x}_{\text{r}})}{ \partial  \boldsymbol{x}_{\text{r}}}$ can be written in the form 
\begin{align}
    \frac{ \partial g(  \boldsymbol{x}_{\text{r}})}{ \partial  \boldsymbol{x}_{\text{r}}}= -\sum_{k=1}^K\frac{P_k(\boldsymbol{x}_{\text{r}})}
    {Q_k(\boldsymbol{x}_{\text{r}})}
\end{align}
where 
\begin{align}
    P_k(\boldsymbol{x}_{\text{r}})&= \frac{2}{\sqrt{\pi}} \ e^{-\pc{\boldsymbol{h}_{1,k} \boldsymbol{x}_{\text{r}}}^2}\boldsymbol{h}_{1,k}+ \frac{2}{\sqrt{\pi}} \ 
    e^{-\pc{\boldsymbol{h}_{2,k} \boldsymbol{x}_{\text{r}}}^2}\boldsymbol{h}_{2,k}
   \\
    Q_k(\boldsymbol{x}_{\text{r}})&= \text{erf}\pc{ \boldsymbol{h}_{1,k} \boldsymbol{x}_{\text{r}} }+ \text{erf}\pc{ \boldsymbol{h}_{2,k} \boldsymbol{x}_{\text{r} } }.
\end{align}
To compute the Hessian $\frac{\partial P_k(\boldsymbol{x}_{\text{r}})}{\partial \boldsymbol{x}_{\text{r}} ^T}$ and $\frac{\partial Q_k(\boldsymbol{x}_{\text{r}})}{\partial \boldsymbol{x}_{\text{r}} ^T}$ are calculated as
\begin{align}
    \frac{\partial P_k(\boldsymbol{x}_{\text{r}})}{\partial \boldsymbol{x}_{\text{r}} ^T}&=-\pc{ \boldsymbol{\boldsymbol{\Psi}}_{1,k}+\boldsymbol{\Psi}_{2,k}}\\
    \frac{\partial Q_k(\boldsymbol{x}_{\text{r}})}{\partial \boldsymbol{x}_{\text{r}} ^T}&=P_k(\boldsymbol{x}_{\text{r}})^T,
\end{align}
where $\boldsymbol{\Psi}_{1,k}$ and $\boldsymbol{\Psi}_{2,k}$ read as
\begin{align}
    \boldsymbol{\Psi}_{1,k}&= \frac{4}{\sqrt{\pi}} e^{-\pc{\boldsymbol{h}_{1,k} \boldsymbol{x}_{\text{r}}}^2} \boldsymbol{h}_{1,k}^T \boldsymbol{x}_{\text{r}}^T \boldsymbol{h}_{1,k}^T \boldsymbol{h}_{1,k}  \\
       \boldsymbol{\Psi}_{2,k}&= \frac{4}{\sqrt{\pi}} e^{-\pc{\boldsymbol{h}_{2,k} \boldsymbol{x}_{\text{r}}}^2} \boldsymbol{h}_{2,k}^T \boldsymbol{x}_{\text{r}}^T \boldsymbol{h}_{2,k}^T \boldsymbol{h}_{2,k} .
\end{align}
The Hessian then reads as
\begin{align}
    \frac{\partial^2 g(\boldsymbol{x}_{\text{r}})}{ \partial  \boldsymbol{x}_{\text{r}}\partial  \boldsymbol{x}_{\text{r}}^T}=
    \displaystyle\sum_{k=1}^K 
    \frac{ P_k(\boldsymbol{x}_{\text{r}})^T P_k(\boldsymbol{x}_{\text{r}})+\pc{\boldsymbol{\Psi}_{1,k}+\boldsymbol{\Psi}_{2,k}} Q_k(\boldsymbol{x}_{\text{r}})}
    {Q_k(\boldsymbol{x}_{\text{r}})^T Q_k(\boldsymbol{x}_{\text{r}})}. 
\end{align}
The Hessian is PSD if $\pc{\boldsymbol{\Psi}_{1,k}+\boldsymbol{\Psi}_{2,k}} Q_k(\boldsymbol{x}_{\text{r}}) \succ 0 \ \forall k \in \chav{1, \hdots K}$. Note that positive semi-definiteness is achieved for $ \boldsymbol{h}_{1,k}\boldsymbol{x}_{\text{r}} \geq 0 \ \forall k \in \chav{1, \hdots K}$ and $ \boldsymbol{h}_{2,k}\boldsymbol{x}_{\text{r}}\geq 0 \ \forall k \in \chav{1, \hdots K}$. Finally, the condition for convexity of the UBMSEP objective function can be cast as 
\begin{align}
    \begin{bmatrix}  \boldsymbol{H}_{\text{R},\theta}^{s^*} - \boldsymbol{H}_{\text{I},\theta}^{s^*} \\ \boldsymbol{H}_{\text{R},\theta}^{s^*} + \boldsymbol{H}_{\text{I},\theta}^{s^*} 
\end{bmatrix} \boldsymbol{x}_\text{r} \geq \boldsymbol{0}.
\end{align}